\documentclass[twocolumn,traditabstract]{aa}
\usepackage{graphicx,amsmath}
\usepackage{epsf}
\usepackage{natbib}
\usepackage{lscape}
\usepackage{epsf}
\usepackage{txfonts}
\usepackage{upgreek}
\usepackage{rotating}
\usepackage{morefloats}
\usepackage{epstopdf}
\usepackage{longtable}

\bibpunct{(}{)}{;}{a}{}{,} 

\def\setsymbol#1#2{\expandafter\def\csname #1\endcsname{#2}}
\def\getsymbol#1{\csname #1\endcsname}

\def\Planck{{\it Planck\/}}

\newbox\tablebox    \newdimen\tablewidth
\def\leaderfil{\leaders\hbox to 5pt{\hss.\hss}\hfil}
\def\tablenote#1 #2\par{\begingroup \parindent=0.8em
    \abovedisplayshortskip=0pt\belowdisplayshortskip=0pt
    \noindent
    $$\hss\vbox{\hsize\tablewidth \hangindent=\parindent \hangafter=1 \noindent
    \hbox to \parindent{\sup{\rm #1}\hss}\strut#2\strut\par}\hss$$
    \endgroup}

\def\L2{\ifmmode L_2\else $L_2$\fi}

\def\DeltaT{\ifmmode \Delta T\else $\Delta T$\fi}
\def\deltat{\ifmmode \Delta t\else $\Delta t$\fi}
\def\fknee{\ifmmode f_{\rm knee}\else $f_{\rm knee}$\fi}
\def\Fmax{\ifmmode F_{\rm max}\else $F_{\rm max}$\fi}
\def\solar{\ifmmode{\rm M}_{\mathord\odot}\else${\rm M}_{\mathord\odot}$\fi}

\def\inv{\ifmmode^{-1}\else$^{-1}$\fi}
\def\mo{\ifmmode^{-1}\else$^{-1}$\fi}
\def\sup#1{\ifmmode ^{\rm #1}\else $^{\rm #1}$\fi}
\def\expo#1{\ifmmode \times 10^{#1}\else $\times 10^{#1}$\fi}
\def\,{\thinspace}
\def\lsim{\mathrel{\raise .4ex\hbox{\rlap{$<$}\lower 1.2ex\hbox{$\sim$}}}}
\def\gsim{\mathrel{\raise .4ex\hbox{\rlap{$>$}\lower 1.2ex\hbox{$\sim$}}}}

\def\simprop{\mathrel{\raise .4ex\hbox{\rlap{$\propto$}\lower 1.2ex\hbox{$\sim$}}}}
\def\deg{\ifmmode^\circ\else$^\circ$\fi}
\def\pdeg{\ifmmode $\setbox0=\hbox{$^{\circ}$}\rlap{\hskip.11\wd0 .}$^{\circ}
          \else \setbox0=\hbox{$^{\circ}$}\rlap{\hskip.11\wd0 .}$^{\circ}$\fi}
\def\arcs{\ifmmode {^{\scriptscriptstyle\prime\prime}}
          \else $^{\scriptscriptstyle\prime\prime}$\fi}
\def\arcm{\ifmmode {^{\scriptscriptstyle\prime}}
          \else $^{\scriptscriptstyle\prime}$\fi}
\newdimen\sa  \newdimen\sb
\def\parcs{\sa=.07em \sb=.03em
     \ifmmode \hbox{\rlap{.}}^{\scriptscriptstyle\prime\kern -\sb\prime}\hbox{\kern -\sa}
     \else \rlap{.}$^{\scriptscriptstyle\prime\kern -\sb\prime}$\kern -\sa\fi}
\def\parcm{\sa=.08em \sb=.03em
     \ifmmode \hbox{\rlap{.}\kern\sa}^{\scriptscriptstyle\prime}\hbox{\kern-\sb}
     \else \rlap{.}\kern\sa$^{\scriptscriptstyle\prime}$\kern-\sb\fi}
\def\ra[#1 #2 #3.#4]{#1\sup{h}#2\sup{m}#3\sup{s}\llap.#4}
\def\dec[#1 #2 #3.#4]{#1\deg#2\arcm#3\arcs\llap.#4}
\def\deco[#1 #2 #3]{#1\deg#2\arcm#3\arcs}
\def\rra[#1 #2]{#1\sup{h}#2\sup{m}}

\def\dots{\relax\ifmmode \ldots\else $\ldots$\fi}
\def\WHzsr{\ifmmode $W\,Hz\mo\,sr\mo$\else W\,Hz\mo\,sr\mo\fi}
\def\mHz{\ifmmode $\,mHz$\else \,mHz\fi}
\def\GHz{\ifmmode $\,GHz$\else \,GHz\fi}
\def\mKs{\ifmmode $\,mK\,s$^{1/2}\else \,mK\,s$^{1/2}$\fi}
\def\muKs{\ifmmode \,\mu$K\,s$^{1/2}\else \,$\mu$K\,s$^{1/2}$\fi}
\def\muKRJs{\ifmmode \,\mu$K$_{\rm RJ}$\,s$^{1/2}\else \,$\mu$K$_{\rm RJ}$\,s$^{1/2}$\fi}
\def\muKHz{\ifmmode \,\mu$K\,Hz$^{-1/2}\else \,$\mu$K\,Hz$^{-1/2}$\fi}
\def\MJysrmK{\ifmmode \,$MJy\,sr\mo$\,mK$_{\rm CMB}\mo\else \,MJy\,sr\mo\,mK$_{\rm CMB}\mo$\fi}
\def\microns{\ifmmode \,\mu$m$\else \,$\mu$m\fi}

\def\muK{\ifmmode \,\mu$K$\else \,$\mu$\hbox{K}\fi}
\def\microK{\ifmmode \,\mu$K$\else \,$\mu$\hbox{K}\fi}
\def\muW{\ifmmode \,\mu$W$\else \,$\mu$\hbox{W}\fi}
\def\kms{\ifmmode $\,km\,s$^{-1}\else \,km\,s$^{-1}$\fi}
\def\kmsMpc{\ifmmode $\,\kms\,Mpc\mo$\else \,\kms\,Mpc\mo\fi}
\setsymbol{LFI:center:frequency:70GHz:units}{70.3\,GHz}
\setsymbol{LFI:center:frequency:44GHz:units}{44.1\,GHz}
\setsymbol{LFI:center:frequency:30GHz:units}{28.5\,GHz}

\setsymbol{LFI:center:frequency:70GHz}{70.3}
\setsymbol{LFI:center:frequency:44GHz}{44.1}
\setsymbol{LFI:center:frequency:30GHz}{28.5}

\setsymbol{LFI:center:frequency:LFI18:Rad:M:units}{71.7\GHz}
\setsymbol{LFI:center:frequency:LFI19:Rad:M:units}{67.5\GHz}
\setsymbol{LFI:center:frequency:LFI20:Rad:M:units}{69.2\GHz}
\setsymbol{LFI:center:frequency:LFI21:Rad:M:units}{70.4\GHz}
\setsymbol{LFI:center:frequency:LFI22:Rad:M:units}{71.5\GHz}
\setsymbol{LFI:center:frequency:LFI23:Rad:M:units}{70.8\GHz}
\setsymbol{LFI:center:frequency:LFI24:Rad:M:units}{44.4\GHz}
\setsymbol{LFI:center:frequency:LFI25:Rad:M:units}{44.0\GHz}
\setsymbol{LFI:center:frequency:LFI26:Rad:M:units}{43.9\GHz}
\setsymbol{LFI:center:frequency:LFI27:Rad:M:units}{28.3\GHz}
\setsymbol{LFI:center:frequency:LFI28:Rad:M:units}{28.8\GHz}
\setsymbol{LFI:center:frequency:LFI18:Rad:S:units}{70.1\GHz}
\setsymbol{LFI:center:frequency:LFI19:Rad:S:units}{69.6\GHz}
\setsymbol{LFI:center:frequency:LFI20:Rad:S:units}{69.5\GHz}
\setsymbol{LFI:center:frequency:LFI21:Rad:S:units}{69.5\GHz}
\setsymbol{LFI:center:frequency:LFI22:Rad:S:units}{72.8\GHz}
\setsymbol{LFI:center:frequency:LFI23:Rad:S:units}{71.3\GHz}
\setsymbol{LFI:center:frequency:LFI24:Rad:S:units}{44.1\GHz}
\setsymbol{LFI:center:frequency:LFI25:Rad:S:units}{44.1\GHz}
\setsymbol{LFI:center:frequency:LFI26:Rad:S:units}{44.1\GHz}
\setsymbol{LFI:center:frequency:LFI27:Rad:S:units}{28.5\GHz}
\setsymbol{LFI:center:frequency:LFI28:Rad:S:units}{28.2\GHz}

\setsymbol{LFI:center:frequency:LFI18:Rad:M}{71.7}
\setsymbol{LFI:center:frequency:LFI19:Rad:M}{67.5}
\setsymbol{LFI:center:frequency:LFI20:Rad:M}{69.2}
\setsymbol{LFI:center:frequency:LFI21:Rad:M}{70.4}
\setsymbol{LFI:center:frequency:LFI22:Rad:M}{71.5}
\setsymbol{LFI:center:frequency:LFI23:Rad:M}{70.8}
\setsymbol{LFI:center:frequency:LFI24:Rad:M}{44.4}
\setsymbol{LFI:center:frequency:LFI25:Rad:M}{44.0}
\setsymbol{LFI:center:frequency:LFI26:Rad:M}{43.9}
\setsymbol{LFI:center:frequency:LFI27:Rad:M}{28.3}
\setsymbol{LFI:center:frequency:LFI28:Rad:M}{28.8}
\setsymbol{LFI:center:frequency:LFI18:Rad:S}{70.1}
\setsymbol{LFI:center:frequency:LFI19:Rad:S}{69.6}
\setsymbol{LFI:center:frequency:LFI20:Rad:S}{69.5}
\setsymbol{LFI:center:frequency:LFI21:Rad:S}{69.5}
\setsymbol{LFI:center:frequency:LFI22:Rad:S}{72.8}
\setsymbol{LFI:center:frequency:LFI23:Rad:S}{71.3}
\setsymbol{LFI:center:frequency:LFI24:Rad:S}{44.1}
\setsymbol{LFI:center:frequency:LFI25:Rad:S}{44.1}
\setsymbol{LFI:center:frequency:LFI26:Rad:S}{44.1}
\setsymbol{LFI:center:frequency:LFI27:Rad:S}{28.5}
\setsymbol{LFI:center:frequency:LFI28:Rad:S}{28.2}

\setsymbol{LFI:white:noise:sensitivity:70GHz:units}{152.6\muKs}
\setsymbol{LFI:white:noise:sensitivity:44GHz:units}{173.1\muKs}
\setsymbol{LFI:white:noise:sensitivity:30GHz:units}{146.8\muKs}

\setsymbol{LFI:white:noise:sensitivity:70GHz}{152.6}
\setsymbol{LFI:white:noise:sensitivity:44GHz}{173.1}
\setsymbol{LFI:white:noise:sensitivity:30GHz}{146.8}

\setsymbol{LFI:white:noise:sensitivity:LFI18:Rad:M:units}{512.0\muKs}
\setsymbol{LFI:white:noise:sensitivity:LFI19:Rad:M:units}{581.4\muKs}
\setsymbol{LFI:white:noise:sensitivity:LFI20:Rad:M:units}{590.8\muKs}
\setsymbol{LFI:white:noise:sensitivity:LFI21:Rad:M:units}{455.2\muKs}
\setsymbol{LFI:white:noise:sensitivity:LFI22:Rad:M:units}{492.0\muKs}
\setsymbol{LFI:white:noise:sensitivity:LFI23:Rad:M:units}{507.7\muKs}
\setsymbol{LFI:white:noise:sensitivity:LFI24:Rad:M:units}{462.2\muKs}
\setsymbol{LFI:white:noise:sensitivity:LFI25:Rad:M:units}{413.6\muKs}
\setsymbol{LFI:white:noise:sensitivity:LFI26:Rad:M:units}{478.6\muKs}
\setsymbol{LFI:white:noise:sensitivity:LFI27:Rad:M:units}{277.7\muKs}
\setsymbol{LFI:white:noise:sensitivity:LFI28:Rad:M:units}{312.3\muKs}
\setsymbol{LFI:white:noise:sensitivity:LFI18:Rad:S:units}{465.7\muKs}
\setsymbol{LFI:white:noise:sensitivity:LFI19:Rad:S:units}{555.6\muKs}
\setsymbol{LFI:white:noise:sensitivity:LFI20:Rad:S:units}{623.2\muKs}
\setsymbol{LFI:white:noise:sensitivity:LFI21:Rad:S:units}{564.1\muKs}
\setsymbol{LFI:white:noise:sensitivity:LFI22:Rad:S:units}{534.4\muKs}
\setsymbol{LFI:white:noise:sensitivity:LFI23:Rad:S:units}{542.4\muKs}
\setsymbol{LFI:white:noise:sensitivity:LFI24:Rad:S:units}{399.2\muKs}
\setsymbol{LFI:white:noise:sensitivity:LFI25:Rad:S:units}{392.6\muKs}
\setsymbol{LFI:white:noise:sensitivity:LFI26:Rad:S:units}{418.6\muKs}
\setsymbol{LFI:white:noise:sensitivity:LFI27:Rad:S:units}{302.9\muKs}
\setsymbol{LFI:white:noise:sensitivity:LFI28:Rad:S:units}{285.3\muKs}

\setsymbol{LFI:white:noise:sensitivity:LFI18:Rad:M}{512.0}
\setsymbol{LFI:white:noise:sensitivity:LFI19:Rad:M}{581.4}
\setsymbol{LFI:white:noise:sensitivity:LFI20:Rad:M}{590.8}
\setsymbol{LFI:white:noise:sensitivity:LFI21:Rad:M}{455.2}
\setsymbol{LFI:white:noise:sensitivity:LFI22:Rad:M}{492.0}
\setsymbol{LFI:white:noise:sensitivity:LFI23:Rad:M}{507.7}
\setsymbol{LFI:white:noise:sensitivity:LFI24:Rad:M}{462.2}
\setsymbol{LFI:white:noise:sensitivity:LFI25:Rad:M}{413.6}
\setsymbol{LFI:white:noise:sensitivity:LFI26:Rad:M}{478.6}
\setsymbol{LFI:white:noise:sensitivity:LFI27:Rad:M}{277.7}
\setsymbol{LFI:white:noise:sensitivity:LFI28:Rad:M}{312.3}
\setsymbol{LFI:white:noise:sensitivity:LFI18:Rad:S}{465.7}
\setsymbol{LFI:white:noise:sensitivity:LFI19:Rad:S}{555.6}
\setsymbol{LFI:white:noise:sensitivity:LFI20:Rad:S}{623.2}
\setsymbol{LFI:white:noise:sensitivity:LFI21:Rad:S}{564.1}
\setsymbol{LFI:white:noise:sensitivity:LFI22:Rad:S}{534.4}
\setsymbol{LFI:white:noise:sensitivity:LFI23:Rad:S}{542.4}
\setsymbol{LFI:white:noise:sensitivity:LFI24:Rad:S}{399.2}
\setsymbol{LFI:white:noise:sensitivity:LFI25:Rad:S}{392.6}
\setsymbol{LFI:white:noise:sensitivity:LFI26:Rad:S}{418.6}
\setsymbol{LFI:white:noise:sensitivity:LFI27:Rad:S}{302.9}
\setsymbol{LFI:white:noise:sensitivity:LFI28:Rad:S}{285.3}

\setsymbol{LFI:knee:frequency:70GHz:units}{29.5\mHz}
\setsymbol{LFI:knee:frequency:44GHz:units}{56.2\mHz}
\setsymbol{LFI:knee:frequency:30GHz:units}{113.7\mHz}

\setsymbol{LFI:knee:frequency:70GHz}{29.5}
\setsymbol{LFI:knee:frequency:44GHz}{56.2}
\setsymbol{LFI:knee:frequency:30GHz}{113.7}

\setsymbol{LFI:knee:frequency:LFI18:Rad:M:units}{16.3\mHz}
\setsymbol{LFI:knee:frequency:LFI19:Rad:M:units}{15.1\mHz}
\setsymbol{LFI:knee:frequency:LFI20:Rad:M:units}{18.7\mHz}
\setsymbol{LFI:knee:frequency:LFI21:Rad:M:units}{37.2\mHz}
\setsymbol{LFI:knee:frequency:LFI22:Rad:M:units}{12.7\mHz}
\setsymbol{LFI:knee:frequency:LFI23:Rad:M:units}{34.6\mHz}
\setsymbol{LFI:knee:frequency:LFI24:Rad:M:units}{46.2\mHz}
\setsymbol{LFI:knee:frequency:LFI25:Rad:M:units}{24.9\mHz}
\setsymbol{LFI:knee:frequency:LFI26:Rad:M:units}{67.6\mHz}
\setsymbol{LFI:knee:frequency:LFI27:Rad:M:units}{187.4\mHz}
\setsymbol{LFI:knee:frequency:LFI28:Rad:M:units}{122.2\mHz}
\setsymbol{LFI:knee:frequency:LFI18:Rad:S:units}{17.7\mHz}
\setsymbol{LFI:knee:frequency:LFI19:Rad:S:units}{22.0\mHz}
\setsymbol{LFI:knee:frequency:LFI20:Rad:S:units}{8.7\mHz}
\setsymbol{LFI:knee:frequency:LFI21:Rad:S:units}{25.9\mHz}
\setsymbol{LFI:knee:frequency:LFI22:Rad:S:units}{15.8\mHz}
\setsymbol{LFI:knee:frequency:LFI23:Rad:S:units}{129.8\mHz}
\setsymbol{LFI:knee:frequency:LFI24:Rad:S:units}{100.9\mHz}
\setsymbol{LFI:knee:frequency:LFI25:Rad:S:units}{38.9\mHz}
\setsymbol{LFI:knee:frequency:LFI26:Rad:S:units}{58.9\mHz}
\setsymbol{LFI:knee:frequency:LFI27:Rad:S:units}{104.4\mHz}
\setsymbol{LFI:knee:frequency:LFI28:Rad:S:units}{40.7\mHz}

\setsymbol{LFI:knee:frequency:LFI18:Rad:M}{16.3}
\setsymbol{LFI:knee:frequency:LFI19:Rad:M}{15.1}
\setsymbol{LFI:knee:frequency:LFI20:Rad:M}{18.7}
\setsymbol{LFI:knee:frequency:LFI21:Rad:M}{37.2}
\setsymbol{LFI:knee:frequency:LFI22:Rad:M}{12.7}
\setsymbol{LFI:knee:frequency:LFI23:Rad:M}{34.6}
\setsymbol{LFI:knee:frequency:LFI24:Rad:M}{46.2}
\setsymbol{LFI:knee:frequency:LFI25:Rad:M}{24.9}
\setsymbol{LFI:knee:frequency:LFI26:Rad:M}{67.6}
\setsymbol{LFI:knee:frequency:LFI27:Rad:M}{187.4}
\setsymbol{LFI:knee:frequency:LFI28:Rad:M}{122.2}
\setsymbol{LFI:knee:frequency:LFI18:Rad:S}{17.7}
\setsymbol{LFI:knee:frequency:LFI19:Rad:S}{22.0}
\setsymbol{LFI:knee:frequency:LFI20:Rad:S}{8.7}
\setsymbol{LFI:knee:frequency:LFI21:Rad:S}{25.9}
\setsymbol{LFI:knee:frequency:LFI22:Rad:S}{15.8}
\setsymbol{LFI:knee:frequency:LFI23:Rad:S}{129.8}
\setsymbol{LFI:knee:frequency:LFI24:Rad:S}{100.9}
\setsymbol{LFI:knee:frequency:LFI25:Rad:S}{38.9}
\setsymbol{LFI:knee:frequency:LFI26:Rad:S}{58.9}
\setsymbol{LFI:knee:frequency:LFI27:Rad:S}{104.4}
\setsymbol{LFI:knee:frequency:LFI28:Rad:S}{40.7}

\setsymbol{LFI:slope:70GHz:units}{$-1.03$\mHz}
\setsymbol{LFI:slope:44GHz:units}{$-0.89$\mHz}
\setsymbol{LFI:slope:30GHz:units}{$-0.87$\mHz}

\setsymbol{LFI:slope:70GHz}{$-1.03$}
\setsymbol{LFI:slope:44GHz}{$-0.89$}
\setsymbol{LFI:slope:30GHz}{$-0.87$}

\setsymbol{LFI:slope:LFI18:Rad:M:units}{$-1.04$\mHz}
\setsymbol{LFI:slope:LFI19:Rad:M:units}{$-1.09$\mHz}
\setsymbol{LFI:slope:LFI20:Rad:M:units}{$-0.69$\mHz}
\setsymbol{LFI:slope:LFI21:Rad:M:units}{$-1.56$\mHz}
\setsymbol{LFI:slope:LFI22:Rad:M:units}{$-1.01$\mHz}
\setsymbol{LFI:slope:LFI23:Rad:M:units}{$-0.96$\mHz}
\setsymbol{LFI:slope:LFI24:Rad:M:units}{$-0.83$\mHz}
\setsymbol{LFI:slope:LFI25:Rad:M:units}{$-0.91$\mHz}
\setsymbol{LFI:slope:LFI26:Rad:M:units}{$-0.95$\mHz}
\setsymbol{LFI:slope:LFI27:Rad:M:units}{$-0.87$\mHz}
\setsymbol{LFI:slope:LFI28:Rad:M:units}{$-0.88$\mHz}
\setsymbol{LFI:slope:LFI18:Rad:S:units}{$-1.15$\mHz}
\setsymbol{LFI:slope:LFI19:Rad:S:units}{$-1.00$\mHz}
\setsymbol{LFI:slope:LFI20:Rad:S:units}{$-0.95$\mHz}
\setsymbol{LFI:slope:LFI21:Rad:S:units}{$-0.92$\mHz}
\setsymbol{LFI:slope:LFI22:Rad:S:units}{$-1.01$\mHz}
\setsymbol{LFI:slope:LFI23:Rad:S:units}{$-0.95$\mHz}
\setsymbol{LFI:slope:LFI24:Rad:S:units}{$-0.73$\mHz}
\setsymbol{LFI:slope:LFI25:Rad:S:units}{$-1.16$\mHz}
\setsymbol{LFI:slope:LFI26:Rad:S:units}{$-0.79$\mHz}
\setsymbol{LFI:slope:LFI27:Rad:S:units}{$-0.82$\mHz}
\setsymbol{LFI:slope:LFI28:Rad:S:units}{$-0.91$\mHz}

\setsymbol{LFI:slope:LFI18:Rad:M}{$-1.04$}
\setsymbol{LFI:slope:LFI19:Rad:M}{$-1.09$}
\setsymbol{LFI:slope:LFI20:Rad:M}{$-0.69$}
\setsymbol{LFI:slope:LFI21:Rad:M}{$-1.56$}
\setsymbol{LFI:slope:LFI22:Rad:M}{$-1.01$}
\setsymbol{LFI:slope:LFI23:Rad:M}{$-0.96$}
\setsymbol{LFI:slope:LFI24:Rad:M}{$-0.83$}
\setsymbol{LFI:slope:LFI25:Rad:M}{$-0.91$}
\setsymbol{LFI:slope:LFI26:Rad:M}{$-0.95$}
\setsymbol{LFI:slope:LFI27:Rad:M}{$-0.87$}
\setsymbol{LFI:slope:LFI28:Rad:M}{$-0.88$}
\setsymbol{LFI:slope:LFI18:Rad:S}{$-1.15$}
\setsymbol{LFI:slope:LFI19:Rad:S}{$-1.00$}
\setsymbol{LFI:slope:LFI20:Rad:S}{$-0.95$}
\setsymbol{LFI:slope:LFI21:Rad:S}{$-0.92$}
\setsymbol{LFI:slope:LFI22:Rad:S}{$-1.01$}
\setsymbol{LFI:slope:LFI23:Rad:S}{$-0.95$}
\setsymbol{LFI:slope:LFI24:Rad:S}{$-0.73$}
\setsymbol{LFI:slope:LFI25:Rad:S}{$-1.16$}
\setsymbol{LFI:slope:LFI26:Rad:S}{$-0.79$}
\setsymbol{LFI:slope:LFI27:Rad:S}{$-0.82$}
\setsymbol{LFI:slope:LFI28:Rad:S}{$-0.91$}

\setsymbol{LFI:FWHM:70GHz:units}{13\parcm01}
\setsymbol{LFI:FWHM:44GHz:units}{27\parcm92}
\setsymbol{LFI:FWHM:30GHz:units}{32\parcm65}

\setsymbol{LFI:FWHM:70GHz}{13.01}
\setsymbol{LFI:FWHM:44GHz}{27.92}
\setsymbol{LFI:FWHM:30GHz}{32.65}

\setsymbol{LFI:FWHM:LFI18:units}{13\parcm39}
\setsymbol{LFI:FWHM:LFI19:units}{13\parcm01}
\setsymbol{LFI:FWHM:LFI20:units}{12\parcm75}
\setsymbol{LFI:FWHM:LFI21:units}{12\parcm74}
\setsymbol{LFI:FWHM:LFI22:units}{12\parcm87}
\setsymbol{LFI:FWHM:LFI23:units}{13\parcm27}
\setsymbol{LFI:FWHM:LFI24:units}{22\parcm98}
\setsymbol{LFI:FWHM:LFI25:units}{30\parcm46}
\setsymbol{LFI:FWHM:LFI26:units}{30\parcm31}
\setsymbol{LFI:FWHM:LFI27:units}{32\parcm65}
\setsymbol{LFI:FWHM:LFI28:units}{32\parcm66}

\setsymbol{LFI:FWHM:LFI18}{13.39}
\setsymbol{LFI:FWHM:LFI19}{13.01}
\setsymbol{LFI:FWHM:LFI20}{12.75}
\setsymbol{LFI:FWHM:LFI21}{12.74}
\setsymbol{LFI:FWHM:LFI22}{12.87}
\setsymbol{LFI:FWHM:LFI23}{13.27}
\setsymbol{LFI:FWHM:LFI24}{22.98}
\setsymbol{LFI:FWHM:LFI25}{30.46}
\setsymbol{LFI:FWHM:LFI26}{30.31}
\setsymbol{LFI:FWHM:LFI27}{32.65}
\setsymbol{LFI:FWHM:LFI28}{32.66}

\setsymbol{LFI:FWHM:uncertainty:LFI18:units}{0.170\arcm}
\setsymbol{LFI:FWHM:uncertainty:LFI19:units}{0.174\arcm}
\setsymbol{LFI:FWHM:uncertainty:LFI20:units}{0.170\arcm}
\setsymbol{LFI:FWHM:uncertainty:LFI21:units}{0.156\arcm}
\setsymbol{LFI:FWHM:uncertainty:LFI22:units}{0.164\arcm}
\setsymbol{LFI:FWHM:uncertainty:LFI23:units}{0.171\arcm}
\setsymbol{LFI:FWHM:uncertainty:LFI24:units}{0.652\arcm}
\setsymbol{LFI:FWHM:uncertainty:LFI25:units}{1.075\arcm}
\setsymbol{LFI:FWHM:uncertainty:LFI26:units}{1.131\arcm}
\setsymbol{LFI:FWHM:uncertainty:LFI27:units}{1.266\arcm}
\setsymbol{LFI:FWHM:uncertainty:LFI28:units}{1.287\arcm}

\setsymbol{LFI:FWHM:uncertainty:LFI18}{0.170}
\setsymbol{LFI:FWHM:uncertainty:LFI19}{0.174}
\setsymbol{LFI:FWHM:uncertainty:LFI20}{0.170}
\setsymbol{LFI:FWHM:uncertainty:LFI21}{0.156}
\setsymbol{LFI:FWHM:uncertainty:LFI22}{0.164}
\setsymbol{LFI:FWHM:uncertainty:LFI23}{0.171}
\setsymbol{LFI:FWHM:uncertainty:LFI24}{0.652}
\setsymbol{LFI:FWHM:uncertainty:LFI25}{1.075}
\setsymbol{LFI:FWHM:uncertainty:LFI26}{1.131}
\setsymbol{LFI:FWHM:uncertainty:LFI27}{1.266}
\setsymbol{LFI:FWHM:uncertainty:LFI28}{1.287}

\setsymbol{HFI:center:frequency:100GHz:units}{100\,GHz}
\setsymbol{HFI:center:frequency:143GHz:units}{143\,GHz}
\setsymbol{HFI:center:frequency:217GHz:units}{217\,GHz}
\setsymbol{HFI:center:frequency:353GHz:units}{353\,GHz}
\setsymbol{HFI:center:frequency:545GHz:units}{550\,GHz}
\setsymbol{HFI:center:frequency:857GHz:units}{857\,GHz}

\setsymbol{HFI:center:frequency:100GHz}{100}
\setsymbol{HFI:center:frequency:143GHz}{143}
\setsymbol{HFI:center:frequency:217GHz}{217}
\setsymbol{HFI:center:frequency:353GHz}{353}
\setsymbol{HFI:center:frequency:545GHz}{550}
\setsymbol{HFI:center:frequency:857GHz}{857}

\setsymbol{HFI:Ndetectors:100GHz}{8}
\setsymbol{HFI:Ndetectors:143GHz}{11}
\setsymbol{HFI:Ndetectors:217GHz}{12}
\setsymbol{HFI:Ndetectors:353GHz}{12}
\setsymbol{HFI:Ndetectors:545GHz}{3}
\setsymbol{HFI:Ndetectors:857GHz}{4}

\setsymbol{HFI:FWHM:Mars:100GHz:units}{9\parcm49}
\setsymbol{HFI:FWHM:Mars:143GHz:units}{7\parcm01}
\setsymbol{HFI:FWHM:Mars:217GHz:units}{4\parcm68}
\setsymbol{HFI:FWHM:Mars:353GHz:units}{4\parcm45}
\setsymbol{HFI:FWHM:Mars:545GHz:units}{4\parcm48}
\setsymbol{HFI:FWHM:Mars:857GHz:units}{4\parcm22}

\setsymbol{HFI:FWHM:Mars:100GHz}{9.49}
\setsymbol{HFI:FWHM:Mars:143GHz}{7.01}
\setsymbol{HFI:FWHM:Mars:217GHz}{4.68}
\setsymbol{HFI:FWHM:Mars:353GHz}{4.45}
\setsymbol{HFI:FWHM:Mars:545GHz}{4.48}
\setsymbol{HFI:FWHM:Mars:857GHz}{4.22}

\setsymbol{HFI:beam:ellipticity:Mars:100GHz}{1.21}
\setsymbol{HFI:beam:ellipticity:Mars:143GHz}{1.03}
\setsymbol{HFI:beam:ellipticity:Mars:217GHz}{1.13}
\setsymbol{HFI:beam:ellipticity:Mars:353GHz}{1.11}
\setsymbol{HFI:beam:ellipticity:Mars:545GHz}{1.12}
\setsymbol{HFI:beam:ellipticity:Mars:857GHz}{1.35}

\setsymbol{HFI:CMB:relative:calibration:100GHz}{$\lsim 1\%$}
\setsymbol{HFI:CMB:relative:calibration:143GHz}{1\%}
\setsymbol{HFI:CMB:relative:calibration:217GHz}{$\lsim 1\%$}
\setsymbol{HFI:CMB:relative:calibration:353GHz}{$\lsim 1\%$}
\setsymbol{HFI:CMB:relative:calibration:545GHz}{}
\setsymbol{HFI:CMB:relative:calibration:857GHz}{}

\setsymbol{HFI:CMB:absolute:calibration:100GHz}{$\lsim 2\%$}
\setsymbol{HFI:CMB:absolute:calibration:143GHz}{$\lsim 2\%$}
\setsymbol{HFI:CMB:absolute:calibration:217GHz}{$\lsim 2\%$}
\setsymbol{HFI:CMB:absolute:calibration:353GHz}{$\lsim 2\%$}
\setsymbol{HFI:CMB:absolute:calibration:545GHz}{}
\setsymbol{HFI:CMB:absolute:calibration:857GHz}{}

\setsymbol{HFI:FIRAS:gain:calibration:accuracy:statistical:100GHz}{}
\setsymbol{HFI:FIRAS:gain:calibration:accuracy:statistical:143GHz}{}
\setsymbol{HFI:FIRAS:gain:calibration:accuracy:statistical:217GHz}{}
\setsymbol{HFI:FIRAS:gain:calibration:accuracy:statistical:353GHz}{2.5\%}
\setsymbol{HFI:FIRAS:gain:calibration:accuracy:statistical:545GHz}{1\%}
\setsymbol{HFI:FIRAS:gain:calibration:accuracy:statistical:857GHz}{0.5\%}

\setsymbol{HFI:FIRAS:gain:calibration:accuracy:systematic:100GHz}{}
\setsymbol{HFI:FIRAS:gain:calibration:accuracy:systematic:143GHz}{}
\setsymbol{HFI:FIRAS:gain:calibration:accuracy:systematic:217GHz}{}
\setsymbol{HFI:FIRAS:gain:calibration:accuracy:systematic:353GHz}{7\%}
\setsymbol{HFI:FIRAS:gain:calibration:accuracy:systematic:545GHz}{7\%}
\setsymbol{HFI:FIRAS:gain:calibration:accuracy:systematic:857GHz}{7\%}

\setsymbol{HFI:FIRAS:zero:point:accuracy:statistical:100GHz}{}
\setsymbol{HFI:FIRAS:zero:point:accuracy:statistical:143GHz}{}
\setsymbol{HFI:FIRAS:zero:point:accuracy:statistical:217GHz}{}
\setsymbol{HFI:FIRAS:zero:point:accuracy:statistical:353GHz}{2.5\%}
\setsymbol{HFI:FIRAS:zero:point:accuracy:statistical:545GHz}{1\%}
\setsymbol{HFI:FIRAS:zero:point:accuracy:statistical:857GHz}{0.5\%}

\setsymbol{HFI:FIRAS:zero:point:accuracy:systematic:100GHz}{}
\setsymbol{HFI:FIRAS:zero:point:accuracy:systematic:143GHz}{}
\setsymbol{HFI:FIRAS:zero:point:accuracy:systematic:217GHz}{}
\setsymbol{HFI:FIRAS:zero:point:accuracy:systematic:353GHz}{2\%}
\setsymbol{HFI:FIRAS:zero:point:accuracy:systematic:545GHz}{2\%}
\setsymbol{HFI:FIRAS:zero:point:accuracy:systematic:857GHz}{2\%}

\setsymbol{HFI:unit:conversion:100GHz:units}{0.2437\MJysrmK}
\setsymbol{HFI:unit:conversion:143GHz:units}{0.3704\MJysrmK}
\setsymbol{HFI:unit:conversion:217GHz:units}{0.4826\MJysrmK}
\setsymbol{HFI:unit:conversion:353GHz:units}{0.2861\MJysrmK}
\setsymbol{HFI:unit:conversion:545GHz:units}{0.05847\MJysrmK}
\setsymbol{HFI:unit:conversion:857GHz:units}{0.002309\MJysrmK}

\setsymbol{HFI:unit:conversion:100GHz}{0.2437}
\setsymbol{HFI:unit:conversion:143GHz}{0.3704}
\setsymbol{HFI:unit:conversion:217GHz}{0.4826}
\setsymbol{HFI:unit:conversion:353GHz}{0.2861}
\setsymbol{HFI:unit:conversion:545GHz}{0.05847}
\setsymbol{HFI:unit:conversion:857GHz}{0.002309}

\setsymbol{HFI:colour:correction:alpha=-2:V2.01:100GHz}{0.9893}
\setsymbol{HFI:colour:correction:alpha=-2:V2.01:143GHz}{0.9759}
\setsymbol{HFI:colour:correction:alpha=-2:V2.01:217GHz}{1.0007}
\setsymbol{HFI:colour:correction:alpha=-2:V2.01:353GHz}{1.0028}
\setsymbol{HFI:colour:correction:alpha=-2:V2.01:545GHz}{1.0019}
\setsymbol{HFI:colour:correction:alpha=-2:V2.01:857GHz}{0.9889}

\setsymbol{HFI:colour:correction:alpha=0:V2.01:100GHz}{1.0008}
\setsymbol{HFI:colour:correction:alpha=0:V2.01:143GHz}{1.0148}
\setsymbol{HFI:colour:correction:alpha=0:V2.01:217GHz}{0.9909}
\setsymbol{HFI:colour:correction:alpha=0:V2.01:353GHz}{0.9888}
\setsymbol{HFI:colour:correction:alpha=0:V2.01:545GHz}{0.9878}
\setsymbol{HFI:colour:correction:alpha=0:V2.01:857GHz}{1.0014}

\def\setsymbol#1#2{\expandafter\def\csname #1\endcsname{#2}}
\def\getsymbol#1{\csname #1\endcsname}
\def\lsim{\mathrel{\raise .4ex\hbox{\rlap{$<$}\lower 1.2ex\hbox{$\sim$}}}}
\def\gsim{\mathrel{\raise .4ex\hbox{\rlap{$>$}\lower 1.2ex\hbox{$\sim$}}}}

\def\simprop{\mathrel{\raise .4ex\hbox{\rlap{$\propto$}\lower 1.2ex\hbox{$\sim$}}}}
\graphicspath{{sed/}{fluxcomp/}{plots/}}

\def\arcm{\ifmmode {^{\scriptscriptstyle\prime}}
          \else $^{\scriptscriptstyle\prime}$\fi}

\def\amin{\ifmmode {^{\scriptscriptstyle\prime}}
          \else $^{\scriptscriptstyle\prime}$\fi}
          
\def\Planck{{\it Planck\/}\ {}}
\def\Planckns{{\it Planck\/}}
\def\Plancks{{\it Planck\/}'s\ {}}
\def\WMAP{{\it WMAP\/}\ {}}
\def\casa{{\sc casa\/}\ {}}

\begin{document}
\raggedbottom 

\title{VLA/JVLA Monitoring of Bright Northern Radio Sources}
   
\author{Noah Kurinsky\inst{\ref{inst1}}
\and Anna Sajina\inst{\ref{inst1}}
\and Bruce Partridge\inst{\ref{inst2}}
\and Steve Myers\inst{\ref{inst3}}
\and Xi Chen\inst{\ref{inst4}}
\and Marcos L\'opez-Caniego \inst{\ref{inst5}}}

\institute{Tufts University, Medford, MA, 02155, USA\label{inst1}
\and Haverford College, Haverford, PA, 19041, USA.\label{inst2}
\and National Radio Astronomy Observatory, Sorocco, NM, 87891, USA\label{inst3}
\and Infrared Processing and Analysis Center, California Institute of Technology, Pasadena, CA 91125, USA\label{inst4}
\and Instituto de F\'\i{sica} de Cantabria (CSIC-Universidad de Cantabria), Avda. los Castros s/n, 39005, Santander, Spain \label{inst5}}

 \abstract
 {We report multiple epoch VLA/JVLA observations of 89 northern hemisphere sources, most with 37\,GHz flux density $>$\,1\,Jy, observed at 4.8, 8.5, 33.5, and 43.3\,GHz. The high frequency selection leads to a predominantly flat spectrum sample, with 85\,\% of our sources being in the \Planck Early Release Compact Source Catalog (ERCSC). These observations allow us to: 1) validate \Plancks 30 and 44\,GHz flux density scale, 2) extend the radio SEDs of \Planck sources to lower frequencies allowing for the full 5-857GHz regime to be studied, and 3) characterize the variability of these sources.  At 30\,GHz and 44\,GHz,  the JVLA and \Planck flux densities agree to within $\sim$\,3\%. On timescales of less than two months the median variability of our sources is 2\%. On timescales of about a year the median variability increases to 14\%. Using the WMAP 7-year data, the 30\,GHz median variability on a 1-6 years timescale is 16\%. }
  
   
 \maketitle

\begin{figure*}[!ht]
\centering
\includegraphics[scale=1.0]{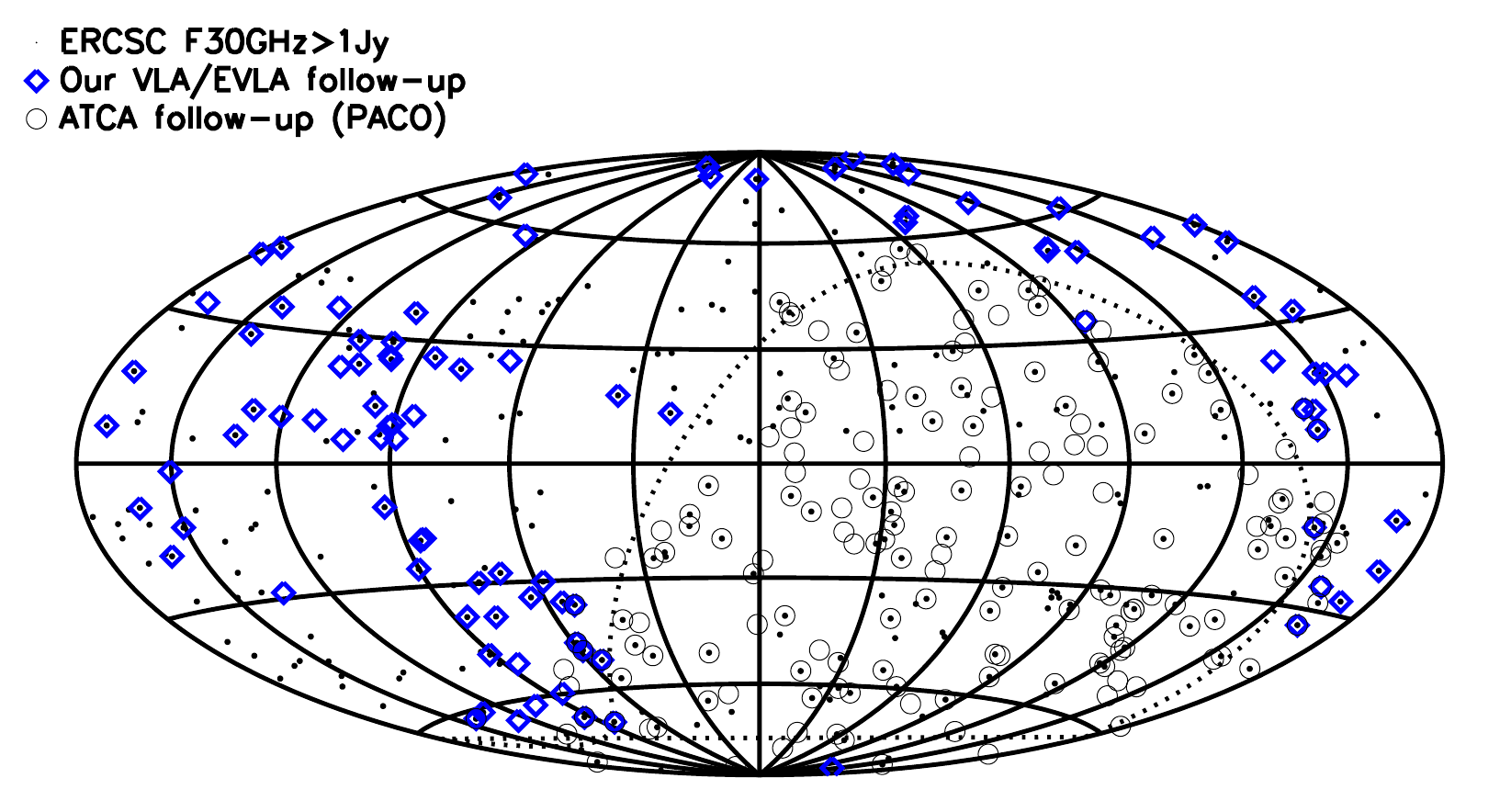}
\caption{Source map for the ERCSC and follow-up studies in Galactic coordinates. The dots mark positions of ERCSC 30\,GHz sources at high Galactic latitude ($|b|>5^\circ$), with a $>1$ Jy cut applied, as at this level the sample is essentially complete.  This is a total of 563 sources. Our study in the northern hemisphere includes 89 sources, the bulk of which are selected to have $S_{37\,GHz}$\,$>$\ ,1\,Jy. For comparison, we plot the $S_{20\,GHz}$\,$>$\,1Jy sources from the AT20G survey \citep{murphy10} in the southern hemisphere. All of these are part of the PACO project, where sources $>$\,500\,mJy make up the PACO bright sample \citep{PACO-proj}. A dotted line separates the sky into the region covered by our study and that covered by the PACO project. Altogether, in the  $>$\,1\,Jy regime, nearly 50\% of the ERCSC 30\,GHz sources have ground-based, near-simultaneous follow-up. }
\label{fig_map}
\end{figure*}

\section{Introduction}\label{sec:introduction} 

The \Planck
satellite \citep{planck_mission}  is capable of all sky observation in nine bands ranging from 30 to 857 GHz.  \Plancks primary goal is to make sensitive observations of temperature and polarization anisotropies in the cosmic microwave background (CMB), but it also offers unprecedented frequency coverage of both Galactic and extragalactic sources, as many of \Plancks bands lie in a frequency range at which accurate observations from the Earth's surface are very difficult due to atmospheric opacity. In addition, \Planck has the ability to observe a given source at all frequencies essentially simultaneously, producing a wealth of spectral data much faster than ground based observatories can.

The first science product of \Planck was the Early Release Compact Source Catalog  \citep[ERCSC;][]{planck_ercsc}, which contains time-averaged flux densities of high reliability sources detected during the first year of \Planck's operation, including 1.6 full sky surveys.The majority of ERCSC sources are brighter than $\sim$\,0.5\,--\,1.0\,Jy, though minimum intensity is also dependent on the frequency of observation. In the low frequency bands considered in this paper, the ERCSC contains hundreds of extragalactic radio sources \citep{planck_ercsc,planck_stats}. These tend to be flat spectrum radio sources, predominantly blazars \citep{planck_blazars,giommi12}. 

The photometric calibration of \Planck is based on the large-scale dipole introduced by the solar motion relative to the cosmic microwave background, an approach unlike the calibration used by ground based observatories. This dipole is very well constrained, leaving a calibration uncertainty of $\sim$\,0.1\%. However, the absolute flux density calibration on very small scales (e.g. for point sources) also depends on a precise knowledge of the \Planck beam at a given frequency, which is less well constrained. Our current understanding of the beam solid angle suggests an uncertainty on the order of 1\%. A cross-comparison of measured \Planck and ground-based absolute flux density scales would therefore help either confirm that the adopted size of the \Planck beam is within the expected uncertainty, or indicate that it may need further refinement. In addition, \Planck provides a potential basis for future absolute flux density scales employed by ground-based facilities such as the Jansky Very Large Array \citep[JVLA, formerly EVLA;][]{perley11_JVLA}. To begin this process, an understanding of the current level of agreement between the two facilities is essential. Such a comparison, however, is complicated by the intrinsic variability of the sources that dominate the catalogs at the lowest \Planck frequencies. This can be mitigated by observing as nearly simultaneously with \Planck as possible, by using larger samples, and by better understanding of the intrinsic variability of the \Planck sources. 

In this paper, we present multiple-epoch VLA/JVLA observations of 89 sources in four spectral bands between 5 and 43\,GHz, most of which are selected to be $>$\,1\,Jy at 37\,GHz, based on observations with the Mets\"ahovi telescope \citep{hovatta2008,hovatta2009}. Preliminary results of this study were presented in \citet{planck_validation}. Based on 32 sources, a VLA-ERCSC flux density comparison found the median 30\,GHz ERCSC flux density to be 8$\pm$4\% brighter than expected based on our VLA observations. Here we extend the previous analysis by including $\sim$\,3$\times$ more sources as well as using updated JVLA/VLA flux density standards. Unlike the case for our earlier paper, here we also have the advantage of multiple epoch observations of  most sources that allow us to characterize the intrinsic variability of the sources over a range of timescales. Our VLA/JVLA observations were all scheduled to be nearly simultaneous with the survey by survey \Planck observations of the same sources, although here we only make comparisons with the time averaged ERCSC data which are publicly available.

Aside from our program, there have been two other programs involving near-simultaneous observations of \Planckns-detected radio sources\footnote{Like our program, these are not true follow up studies of \Planckns-detected sources, which cannot be done simultaneously with \Planckns's observations, since \Planck data reduction is time consuming. Instead, all three programs used their knowledge of \Plancks scanning strategy combined with known samples of bright radio sources to produce samples likely to heavily overlap with detections made by \Planck}.  The first of these programs was the PACO project, conducted in the southern hemisphere \citep{PACO-proj}, which observed 20\,GHz-selected sources using the Australia Telescope Compact Array (ATCA).  Sources were observed essentially simultaneously at 5.5, 9, 18, 21, 33 and 39\,GHz. The second of these programs was SiMPlE, which followed-up 250 5\,GHz-selected sources in the northern hemisphere using the single dish Medicina telescope at 5 and 8\,GHz as well as at 21\,GHz, with observations made between Aug. 2010 and Aug. 2011 \citep{procopio11}. Compared to these surveys, our work covers a smaller number of sources (89); however, our predominantly 37\,GHz selection is most sensitive to the flat spectrum sources that dominate the \Planck-detected radio sources.  We use the JVLA whose Ka and Q-bands (33.45 and 43.22\,GHz respectively) closely match \Planckns's 28.46 and 44.10\,GHz bands. These characteristics make our sample particularly well suited to the flux density comparison presented in this paper. The SiMPlE study is the only one using a single dish radio telescope, and hence has the advantage of providing accurate flux densities for extended sources. While all of these programs have their strengths and weaknesses, the intrinsic variability of the \Planck radio sources means that ultimately the more such ancillary data is collected close in time to \Plancks's observations of the same sources, the more value is added to the eventual \Planck legacy point source catalog. Thanks to these three surveys, the vast majority of the brightest radio sources at $\sim$\,5-40\,GHz now have multiple epoch, near-simultaneous, ground-based observations. These observations extend the SEDs of the \Planck detected blazars and flat spectrum radio galaxies to lower frequencies, and add data on the variability of these sources. 

Our paper is organized as follows. In Section\,\ref{sec:data}, we present the sample selection, observations, and data reduction. In Section\,\ref{sec:match} we present the cross-matching of our sample with the ERCSC. In Section\,\ref{sec:seds}, we analyze the SED types of our sources. In Section\,\ref{sec:variability}, we characterize their variability. In this section, we also make use of data from the  Wilkinson Microwave Anisotropy Probe \citep[WMAP; ][]{bennett03} to provide a longer timescale for our variability analysis. Finally, in Section\,\ref{sec:fluxcomp}, we compare the \Planck and JVLA flux densities at 30 and 44\,GHz. The summary and conclusions are presented in Section\,\ref{sec:summary}. 

\section{Data \label{sec:data}}

\subsection{VLA/JVLA Sample \label{sec:sample}}

Our main goal in selecting galaxies for our survey was to observe as many sources that would be bright enough at the 30\,--\,40\,GHz frequency range to have a high probability of being detected by the \Planck Low Frequency Instrument \citep[LFI;][]{planck_lfi}, which operates at frequencies of 30, 44, and 70\,GHz. Our specific choice of VLA/JVLA observing targets was dictated by the goal of observing them as  close to simultaneous with the \Planck observations as possible.  

The first results of our study presented in \citet{planck_validation} included 32 sources chosen from the VLA calibrator list, where we specifically looked for 5 or 8\,GHz flux densities $>$\,1Jy. These published sources represent only the subset of the ones observed that were matched in the ERCSC at 30\,GHz. In the present paper, we exclude the observations taken on July 24, 2009 since, for that day, we could not obtain primary flux calibrator measurements, but used a less reliable secondary calibrator. This leaves us with 23 of the sources in \citet{planck_validation}. We further include here data obtained on 16 sources in the same program on Nov. 15 and 18th, 2009, as well as on 11 more VLA sources that were not published in \citet{planck_validation} because they did not have ERCSC counterparts.

In the present paper, we expand the sample with new JVLA data for 70 sources, which includes repeated observation of some of the above mentioned VLA sources. Our JVLA targets  were predominantly selected from the Complete Northern 1Jy Sample, which includes all 104 sources above declination -10$^{\circ}$ whose average 37\,GHz flux density is $>$\,1\,Jy, as determined through long-term 37\,GHz monitoring by the Metsahovi telescope \citep[e.g.][]{hovatta2008,hovatta2009,planck_blazars}\footnote{However, in instances where there were insufficient numbers of 37\,GHz-selected sources to fill the required  1\,hour scheduling blocks (see Section\,\ref{sec:observations}) sources selected as above were used to fill in the block.}.  In this paper, while the analysis focuses on the sources with ERCSC counterparts, we publish the data for all sources we observed near simultaneously with \Planck for the sake of upcoming \Planck catalogs which will reach lower flux densities than the ERCSC. We still exclude from the analysis three sources for which we measure 33\,GHz flux densities $<$\,300\,mJy (these are cases where the wrong source was observed, or cases where the bulk of the flux is resolved out in our relatively high resolution imaging). 

In summary, the sample we analyze includes 89 sources with a total of 159 separate observations. While the above makes it clear that the sample is neither homogeneous nor complete, 66 of our sources ($\sim$\,75\,\% of the total) are among the 104 in the Complete Northern 1\,Jy Sample. These sources are therefore dominant, and largely determine the properties of our sample discussed in the rest of the paper. The spatial distribution of our sources can be seen in Figure \ref{fig_map}. 

\subsection{Observations}
\label{sec:observations}

Our observations were made in two intervals of time, separated by approximately 6 months during which the NRAO array was down during the conversion from the VLA to the JVLA. In our earlier VLA observations, we used the K band detectors, centered at 22.46 GHz, whereas after the conversion to the JVLA, we switched to the Ka band (33.45\,GHz). The Ka band is preferable to K band, because it is closer to \Plancks lowest frequency of 28.46\,GHz, and also less affected by the atmospheric water vapor line at 22\,GHz.  

Observations were made in 23 observing blocks (Table \ref{sb_table})\footnote{For a few sources in our sample, we included earlier observations from \citet{sajina11} and, in the case of J0555+3948, earlier NRAO calibration observations (www.aoc.nrao.edu/$\sim$smyers/JVLApolcal/polcal\_master.html).}. In the case of the JVLA observations, these were typically 1 hour blocks. Each observing block included one of three standard calibrators (Table \ref{cal_table}),  as well as checks on array pointing, and 7-9 science targets. It is important to keep in mind that the bulk of these observations were done during very early stages of the JVLA's commissioning, when instrumental problems were to be anticipated. Given the brightness of our sources, even in the worst cases, the statistical error is below 0.5\% (see Section \ref{datred} for details on observational errors and data flagging).

In all cases, the choice of specific targets to observe depended most strongly on the parts of the sky that the \Planck satellite was scanning at the time of observation. With the queue-based system for JVLA scheduling, and given our higher demand for good weather, perfect simultaneity was not feasible. However, in the vast majority of cases, observations by \Planck and the JVLA are separated by at most two weeks. 

All of our observations were made with the VLA/JVLA array in its more compact configurations --  D, C or DnC -- where the angular resolution (the half power beam width) is 1-3\arcsec at K band, 0.6\,-\,2\arcsec at Ka band and 0.5\,--\,1.5\arcsec at Q band\footnote{http://JVLAguides.nrao.edu/index.php?title=Observational\_Status\_ Summary\_-\_Current\#Performance\_of\_the\_JVLA} (22.46, 33.45, and 43.22\,GHz respectively). With the exception of J1230+1223 (i.e. M87), our sources were unresolved even at the highest frequency in the highest resolution configuration, so we can generally compare flux density measurements made in different configurations. In comparison, the \Planck beam widths, as adopted in the ERCSC are 32.65\arcmin at 30\,GHz and 27.00\arcmin at 44\,GHz \citep{planck_ercsc}. 

\subsection{Data Reduction}\label{datred}

Data reduction of the JVLA observations was done using the \casa software package\footnote{www.casa.nrao.edu}. We first performed all the standard initial reduction steps, including correction for atmospheric opacity, antenna delay solutions, and bandpass corrections. In addition, 15-20\% of the raw data had to be removed before further analysis due to errors related to receiver malfunction, temporarily de-comissioned antennae (due to the ongoing transition from VLA to JVLA), or time lapses in antenna slewing. Each source was amplitude and phase calibrated as well.

In each block, we included observations of one of three primary flux calibrators (Table \ref{cal_table}), which provided the absolute flux density scale. We adopted the Perley-Butler 2010 absolute flux calibration standards. The VLA data originally presented in \citet{planck_validation} were calibrated using the older Perley-Taylor 1999 standards, which we rescaled to the Perley-Butler 2010 flux standards. Table\,\ref{cal_table} gives the adopted calibrator flux densities, as well as the correction factors applied to the earlier VLA data. 

After calibration, an image was produced for each observation. Each target source was fitted with a two dimensional gaussian providing peak and integrated flux densities as well as their associated errors. We checked for extended sources by looking for discrepancies between peak and integrated fluxes and by examination of images. We found one extended source, J1230+1223 (i.e. M87, 3C274). In Table \ref{table_reddata} we present the integrated flux densities and flux density errors for each source, observation, and band. The flux density errors listed represent the quadrature sum of the errors from the above fitting, which are dominated by the instrumental noise in the images, and calibration uncertainties which are determined by the standard deviation of the flux densities of 3C48 and 3C286 (the primary flux calibrators) at each frequency.  At the Ka and Q bands, the uncertainty introduced by the scatter in the measurements of the calibrators is estimated at $\sim$0.1\,--\,0.2\,\,\%.  

\subsection{The Early Release Compact Source Catalog \label{sec:planckdata}}
In this paper, we obtain \Planck flux densities for our sources from the \Planck Early Release Compact Source Catalog \citep[ERCSC;][]{planck_ercsc}. 
The ERCSC contains high reliability ($>$\,90\% cumulative reliability) compact sources (both Galactic and extragalactic) based on \Plancks first year of observations. In this paper, we adopt the standard flux density and flux density error estimates, as given in the {\sc flux} and {\sc flux\_err} columns in the ERCSC catalogs. These are estimated through aperture photometry with a radius equal to the sky-averaged FWHM for the given band, with aperture corrections applied \citep{planck_ercsc}. The ERCSC also includes flux density estimates based on fitting a 2D gaussian  ({\sc gauflux}) as well as fitting the point-spread function {\sc psfflux}. The {\sc flux} column is generally preferred since the {\sc gauflux} measurements are more sensitive to the presence of any underlying extended emission, while the {\sc psfflux} measure is affected by the fact that the exact shape of the \Planck\ beam is not quite constant across the sky. In all cases, these flux densities are in fact averages of all data obtained for a given source by \Planck observations at a given frequency. Since the ERCSC contains about 1.6 full sky surveys, the bulk of the sources have fluxes averaged over 2 or more time periods (with the sources closest to the ecliptic poles having the most frequent observations).

We note explicitly that the 44 GHz \Planck measurements were generally noisier than those at 30\,GHz. In addition, as discussed in \citet{planck_validation}, the placement of the three 44 GHz horns in the \Planck focal plane ensured that there is an approximately one week separation in time between observations of a given source by one horn and the other two horns at that frequency. Thus variability is likely to play a larger role at 44 GHz than at  30 GHz, where the horn separation is smaller. 


\begin{figure*}[!ht]
\includegraphics[scale=0.25,clip=true,trim=25 15 20 28]{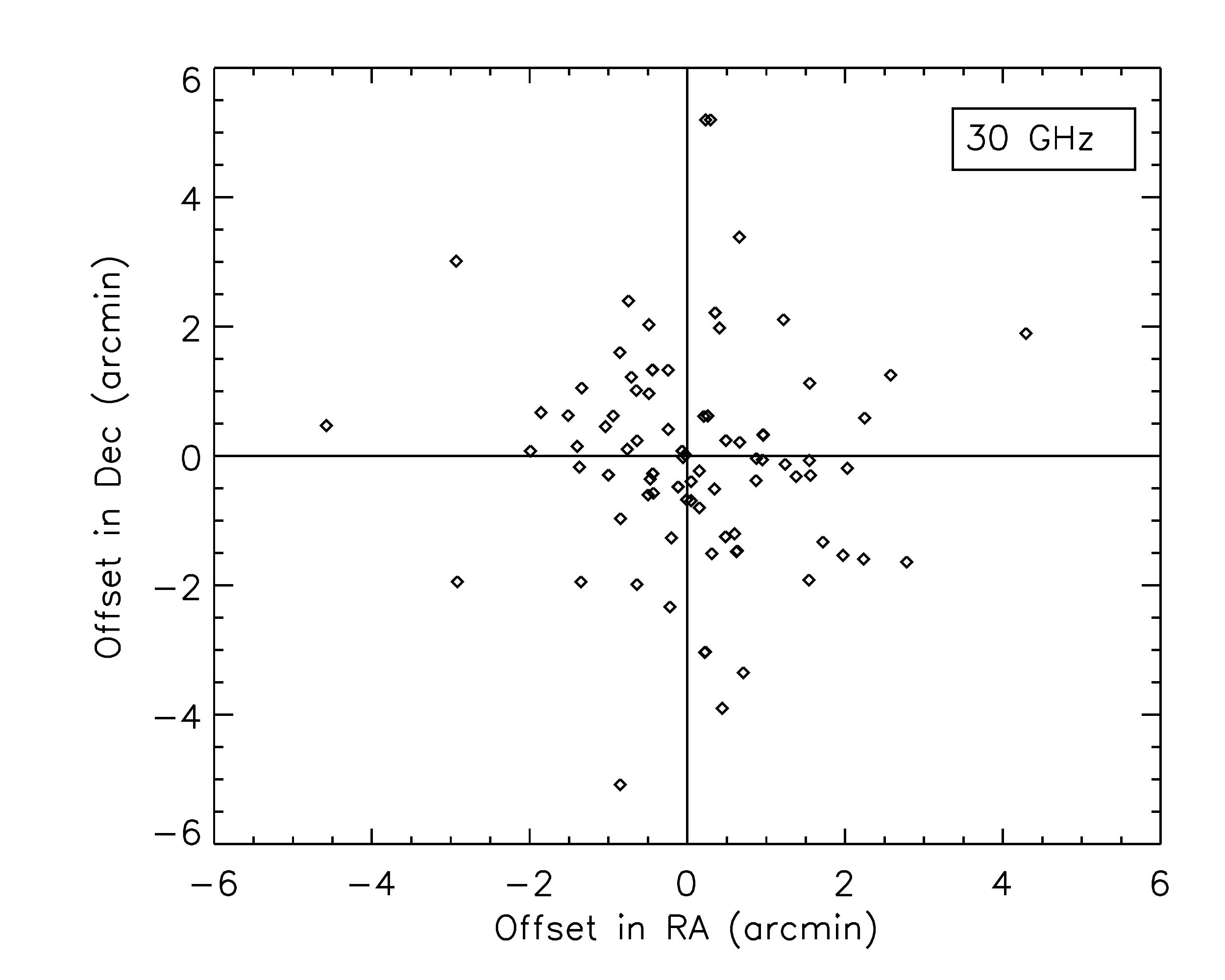}
\includegraphics[scale=0.25,clip=true,trim=25 15 20 28]{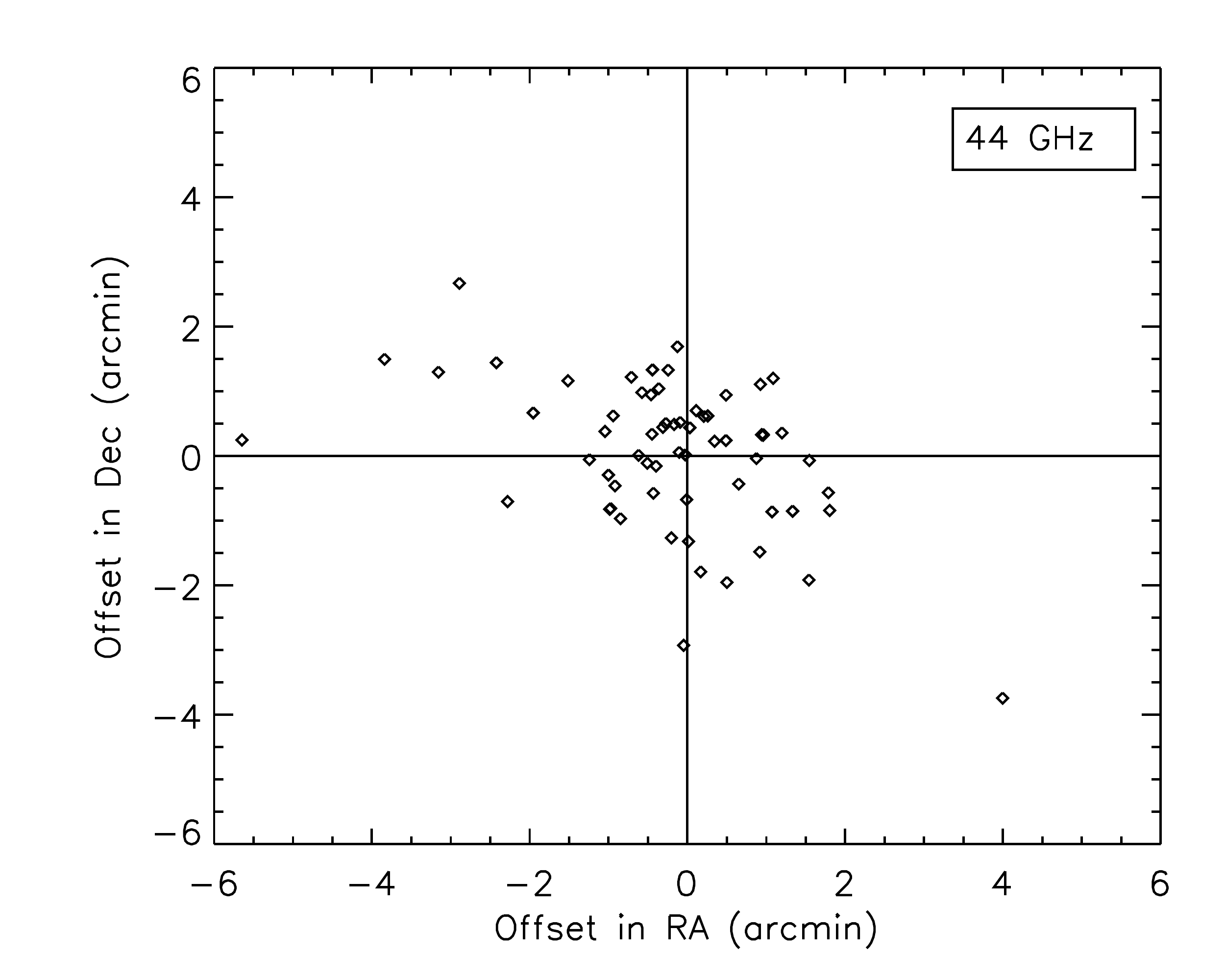}
\includegraphics[scale=0.25,clip=true, trim=25 15 20 28]{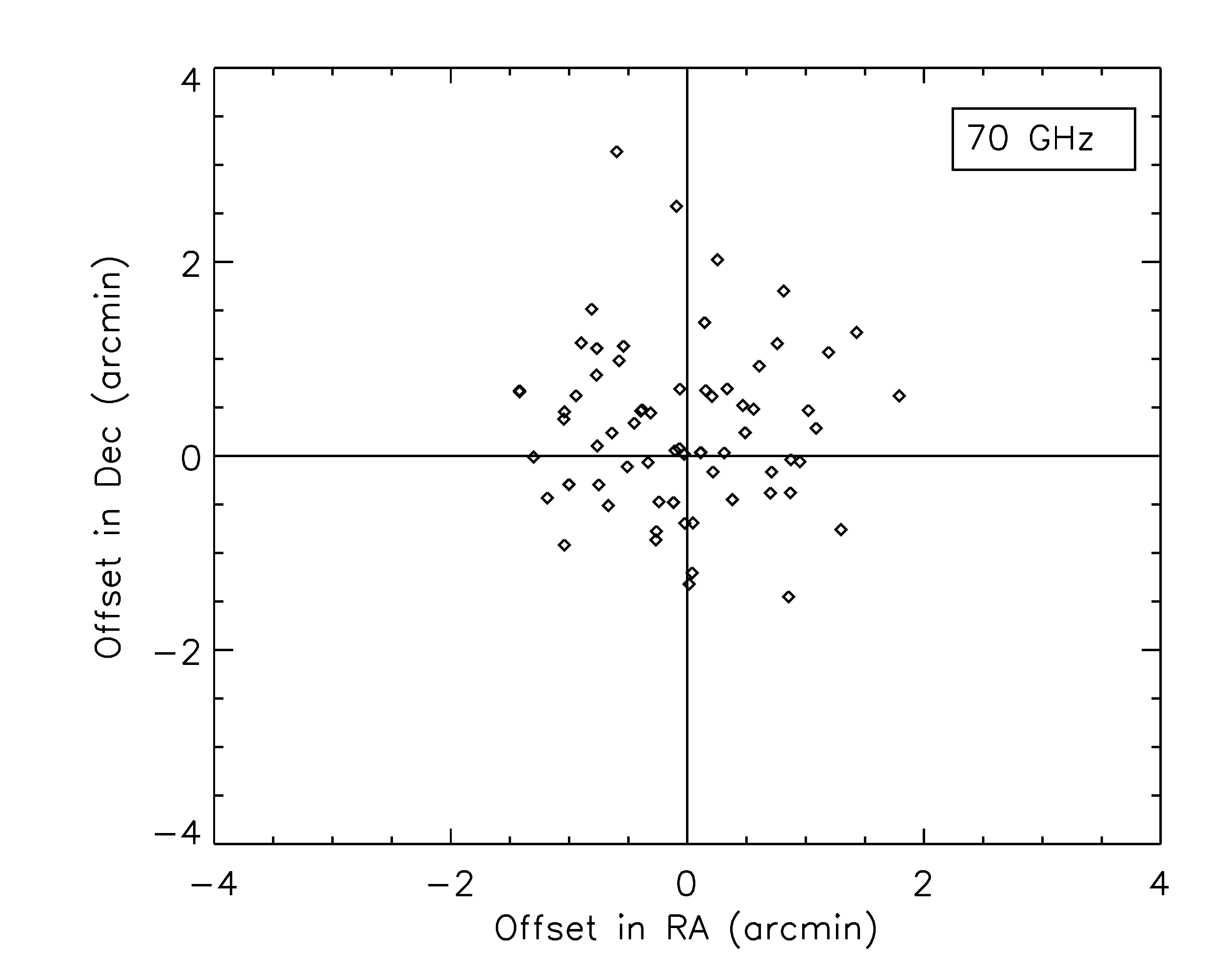}
\caption{Positional scatter between the Planck sources at 30, 44\,GHz, and 70\,GHz and their JVLA matches. For the 70\,GHz comparison, the JVLA Ka positions are used.}
\label{position_scat}%
\end{figure*}

\begin{figure*}[!ht]
\centering
\includegraphics[scale=0.46,clip=true, trim=25 15 20 34]{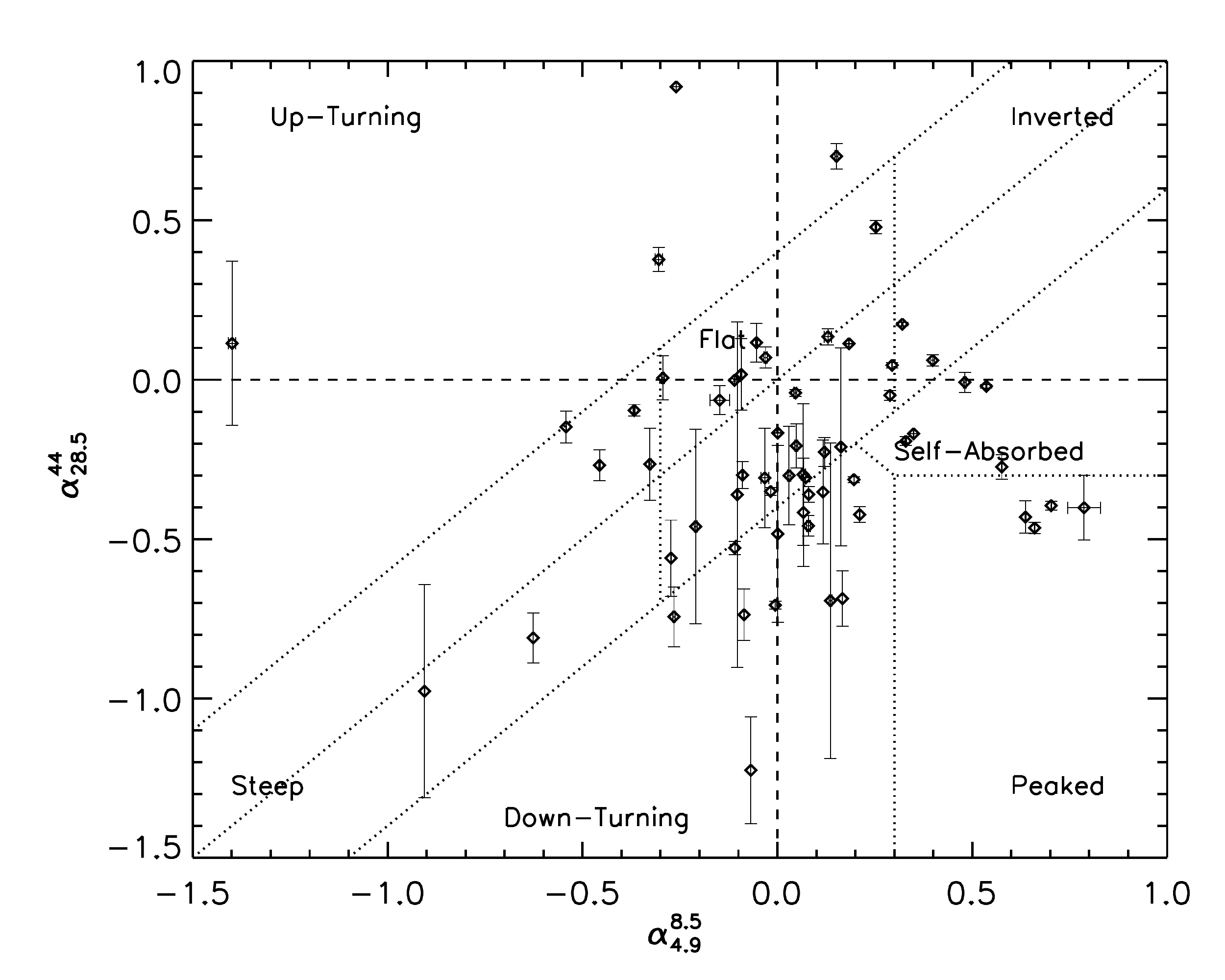}
\caption{The 5-40\,GHz radio source color-color diagnostic plot. Note that here we extrapolate both the VLA K-band and the JVLA Ka-band observations to 28.46\,GHz, the center frequency of the \Planck 30\,GHz band. All observations for a given source are averaged, such that each data point represents a unique source. For ease of comparison, we adopt the \citet{PACO-proj} definitions of "flat", "steep", "up-turning", "inverted", "self-absorbed" and "peaked" spectra (as defined by the dotted regions).}
\label{color_color}
\end{figure*}


\subsection{The \WMAP 7-Year Data}\label{sec:w7}

The \WMAP data used in this paper comes from the \WMAP Seven Year release \citep{wmap_7a,wmap_7b}. The 7-year catalog contains 471 sources which represent 5\,$\sigma$ peaks in the maps co-adding all data from the first 7 years of the survey in each of the 5 WMAP bands. The flux densities in this catalog are the mean flux densities of the sources across these 7 years. 
WMAP also produced single year maps, where (like the \Planck observations) a given source has typically been observed more than once in the span of that year. Therefore even flux densities based on these single year maps are averaged over multiple observations. We obtain such yearly-averaged flux densities for the 471 sources in the 7-year catalog through the point source variability table\footnote{\tiny{http://lambda.gfsc.nasa.gov/product/map/dr4/ptsrc\_variability\_info.cfm}}. Here we only make use of the WMAP Ka (30.0\,GHz) data.



\begin{figure*}[!ht]
\centering
\includegraphics[scale=0.4]{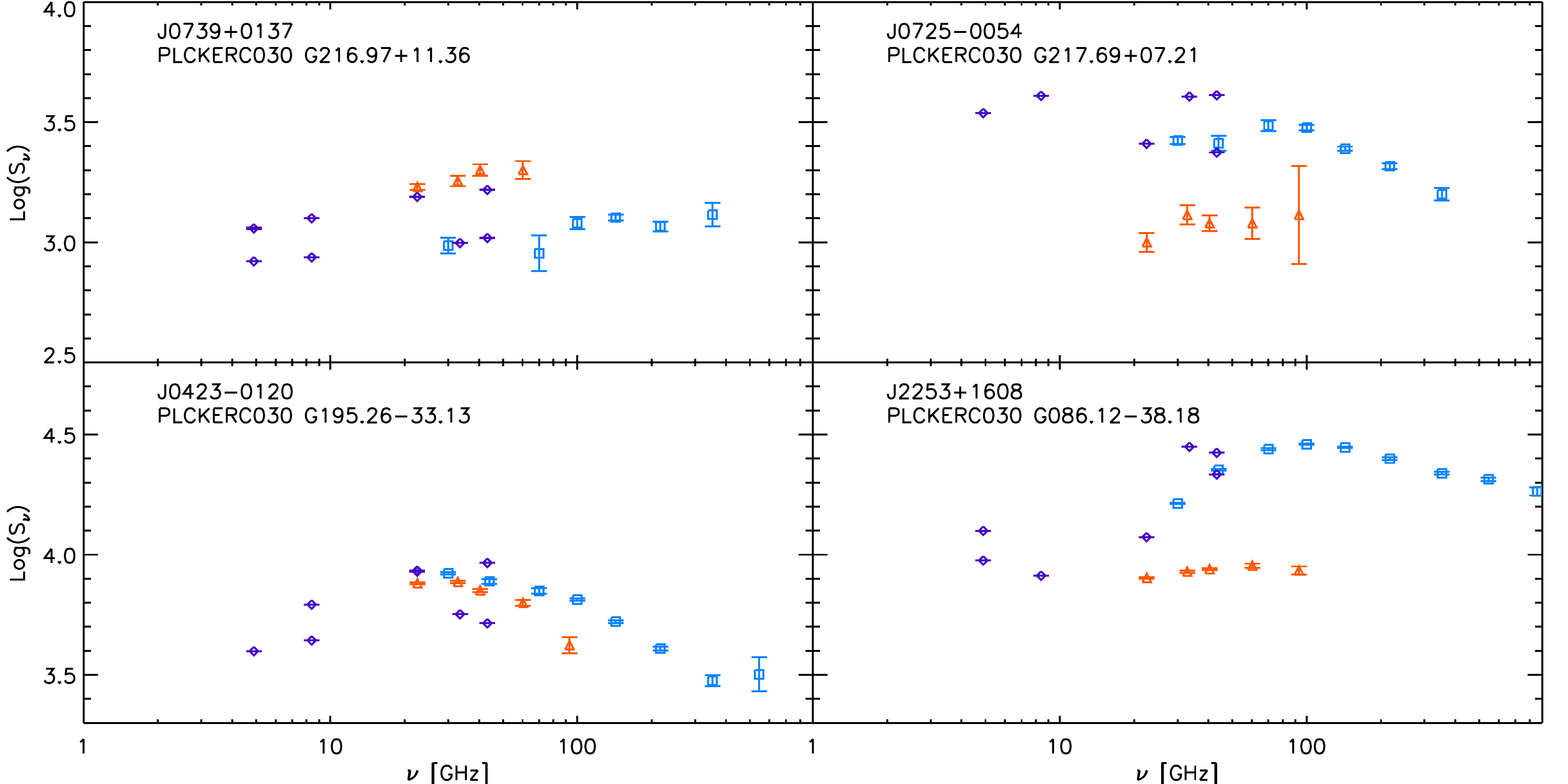}
\caption{The four most highly variable sources in our sample. The purple diamonds indicate VLA/JVLA data, orange triangles indicate the 7-year \WMAP average flux densities of the sources, and light blue squares indicate the \Planck ERCSC flux densities. \label{vars}}
\end{figure*}


\begin{figure*}[!ht]
\centering
\includegraphics[width=\textwidth,clip=true]{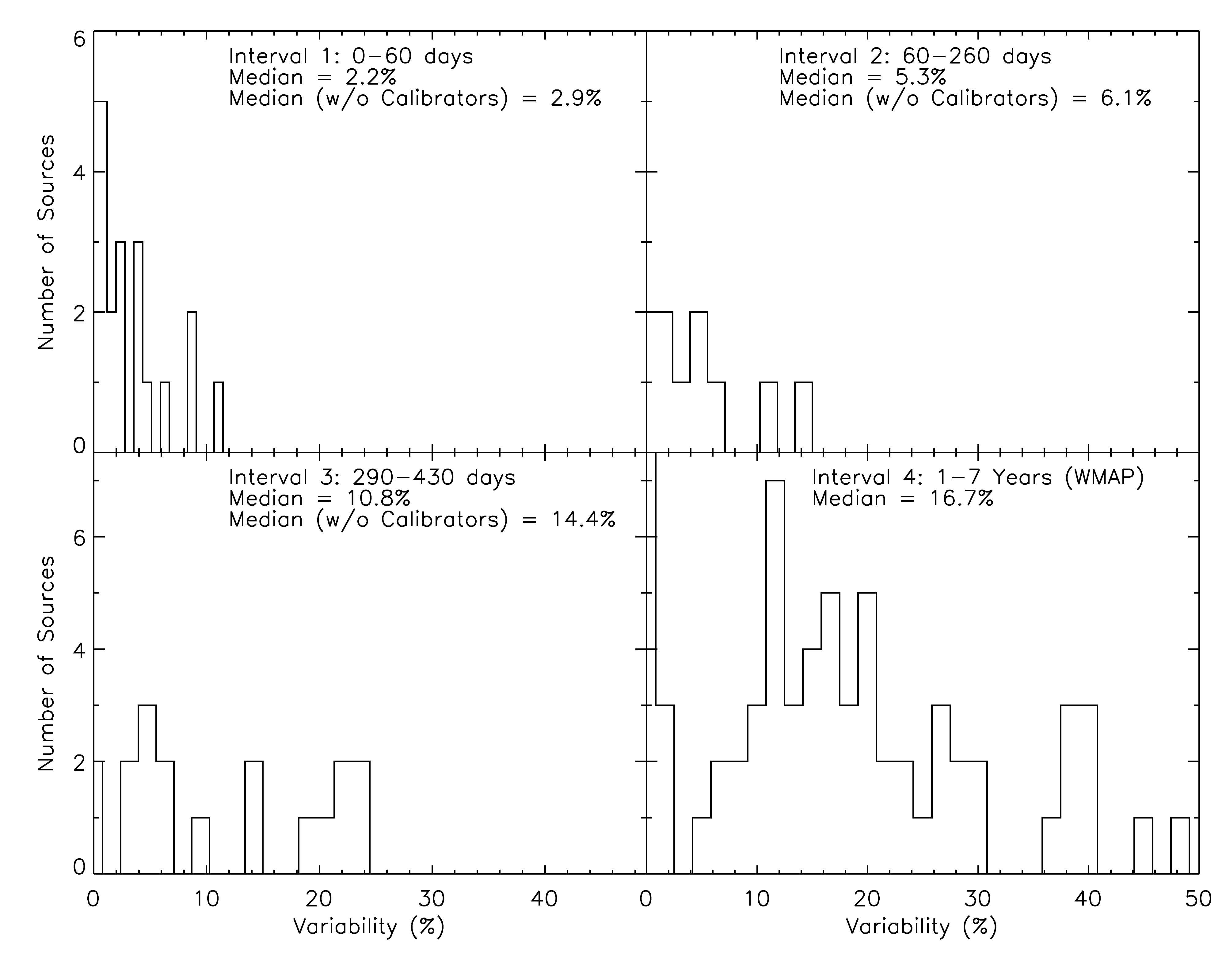}
\caption{Variability distribution on different timescales. The first three time intervals are based on our VLA/JVLA data alone. The last interval (1-7 years) is based on the yearly-averaged  WMAP data (see text for details). All variability indices are estimated for the JVLA Ka band (33.5\,GHz), or 33.0\,GHz for WMAP.}
\label{fig_variability}
\end{figure*}


\section{Results}

\subsection{Source Matching}\label{sec:match}

We consider a given JVLA-ERCSC source pair a match if the distance between them is less than 0.75\,FWHM of the \Planck beam for a given band. As seen in Figure\,\ref{position_scat}, the positional offsets for our matches never exceed $\sim$\,6\,arcmin and are typically within $\sim$\,2\,arcmin. Given the scarcity of extragalactic radio sources at the 1\,Jy level, the chance of random association is negligible. However, in one case, the nominal \Planck source is located in-between two of our sources (both listed as J1310+3220 in Table\,\ref{table_reddata}), and is the sum of the two. These sources are omitted from further analysis.

Of our 89 sources, we found 76 ERCSC matches at 30\,GHz, and a subset of these (59) have ERCSC matches at 44\,GHz (listed in Table\,\ref{table_matches}). The typically negative 30\,--\,44\,GHz spectral indices of our sources explain the lower number of matches at 44\,GHz. This is also the result of the lower sensitivity of the \Planck\  44\,GHz band, as mentioned in Section\,\ref{sec:planckdata}. 

\subsection{Sample Characterization and SED Types}\label{sec:seds}

Figure\,\ref{color_color} shows a diagnostic color-color plot, providing an overview of the SEDs of our sources.  It is obvious that our sample is biased toward flat-spectrum sources. In Table\,\ref{table_classes}, we give the fractional distribution of each of the classes shown in Figure\,\ref{color_color}, as well as the fractional composition of the PACO Bright and Faint samples \citep{PACO-proj,PACO-proj-faint}.   We find that our sample has a much higher concentration of flat spectrum sources, compared with both PACO samples. This is unsurprising, given that the PACO selection is at  20\,GHz, whereas our sample's primary selection is at 37\,GHz. Naturally, the higher the frequency used to select a sample, the higher the fraction of flat spectrum relative to steep spectrum sources \citep[e.g.][]{sajina11}. In addition, the PACO bright sample is based on $S_{20GHz}$\,$>$\,0.5\,Jy, whereas ours has a threshold of $S_{37\,GHz}$\,$>$\,1Jy (for the $\sim$\,70\% of the sample that are in the 1\,Jy Complete Northern Sample). Adopting a typical spectral index\footnote{Defined as $S \propto \nu^\alpha$.} of -0.4, our flux density limit translates to 1.3\,Jy at 20\,GHz, brighter than the PACO bright sample limit.  

As we demonstrate in Section\,\ref{sec:variability}, our sample shows significant variability. Four of the most highly variable sources are shown in Figure\,\ref{vars}, as an illustration. This variability means that a given source can be classified differently depending on when it is observed (both J0423-0120 and J2253+1608 clearly change the sign of their $\alpha_{28.5}^{44}$ from one observed epoch to the next). For simplicity, Figure\,\ref{color_color} is based on the average flux densities of our sources at each of the frequencies shown. 

\subsection{Variability Analysis \label{sec:variability}}

The availability of multi-epoch observations allows us to study the variability of our sources, which is expected to be significant given that the bulk of our sample are flat spectrum sources, most likely blazars. For the variability analysis, we adopt the variability index formula from \citet{sadler06} which we reproduce in Equation\,\ref{eq_var}. In our implementation, we calculate the variability indices from pairs of observations, therefore N\,=\,2 in Equation\,\ref{eq_var}. The variability is assessed on the basis of the Ka (33\,GHz) flux densities from multiple observations of the same source.  The available K (22\,GHz) flux densities are included in the analysis, after being interpolated to the Ka central frequency using the K-Q spectral indices.

\begin{equation}
V_{rms}=\frac{100}{\langle S\rangle}\sqrt{\frac{\sum(S_{i}-\langle S\rangle)^2-\sum\sigma_{i}^2}{N}}
\label{eq_var}
\end{equation}

\begin{figure*}[!ht]
\centering
\includegraphics[scale=0.38,clip=true,trim=25 15 16 34]{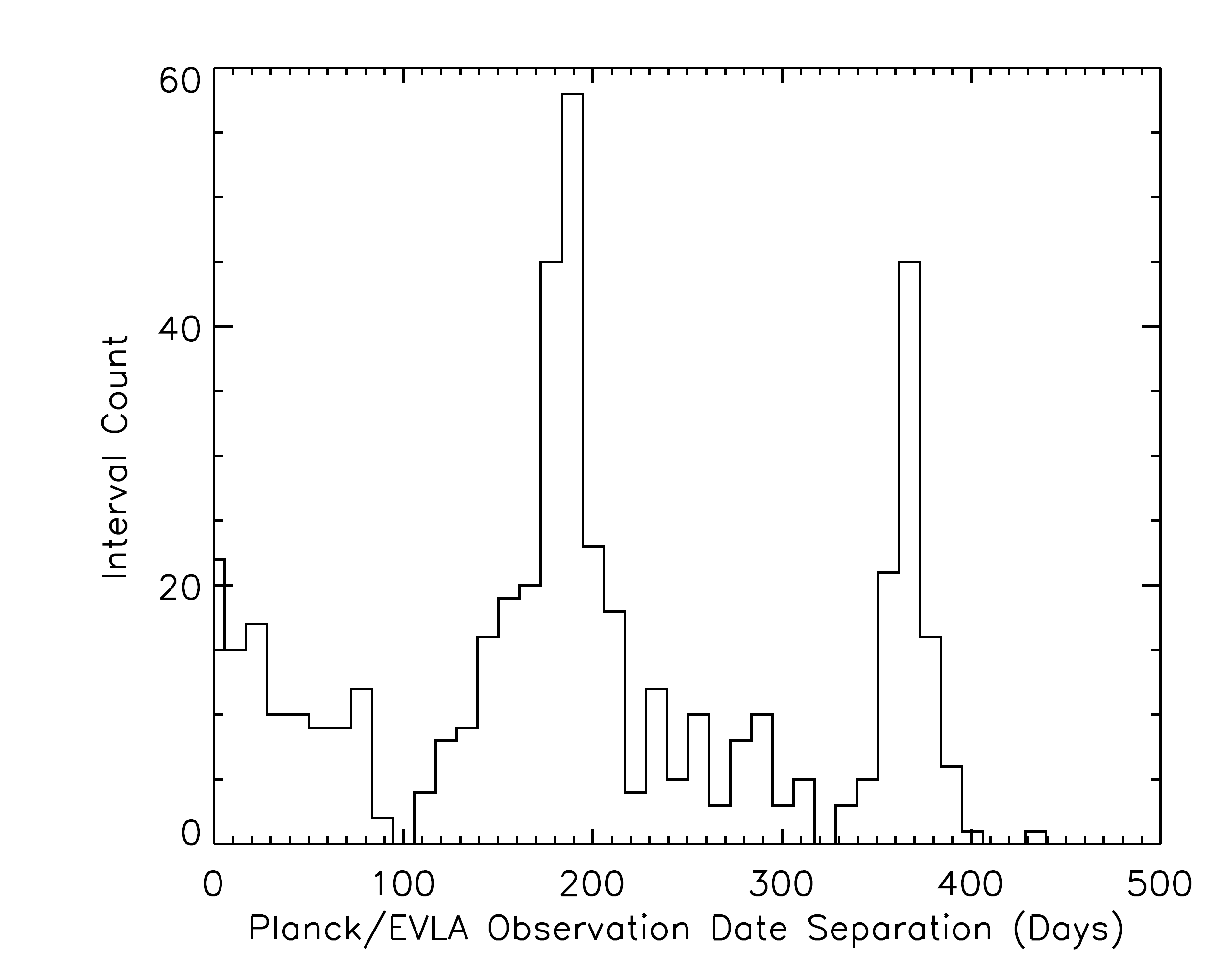}
\caption{Time intervals between our JVLA observations of a given source and the associated observations in the ERCSC. It is important to note that the bulk of our JVLA observations were actually taken after the end of period included in the ERCSC, but were timed so that in the majority of cases they were within a few weeks of the later \Planck observations of the same source. }
\label{fig_obsvar}
\end{figure*}

\begin{figure*}[!ht]
\centering
\includegraphics[scale=0.40,clip=true,trim=50 15 25 34]{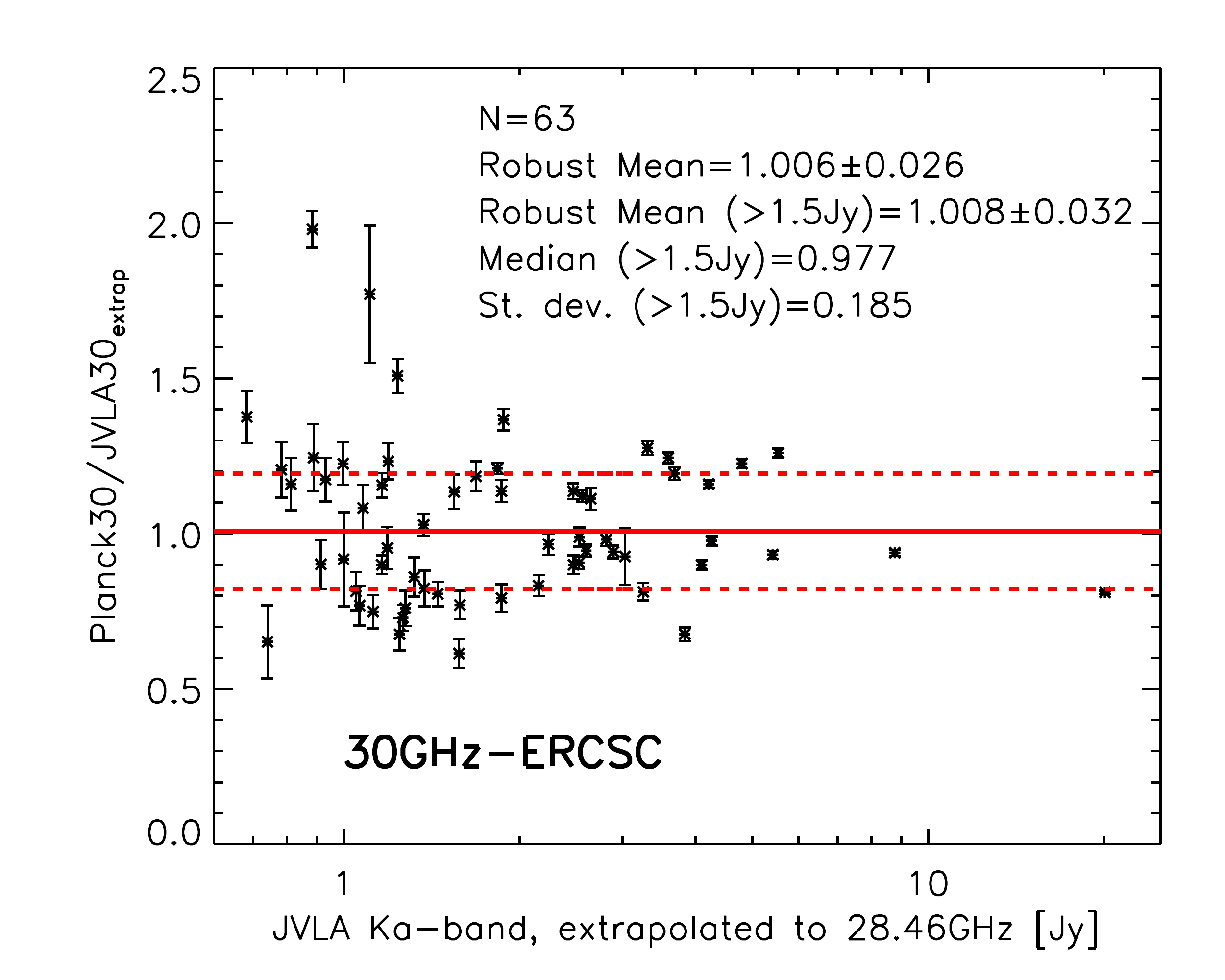}
\includegraphics[scale=0.40,clip=true,trim=50 15 25 34]{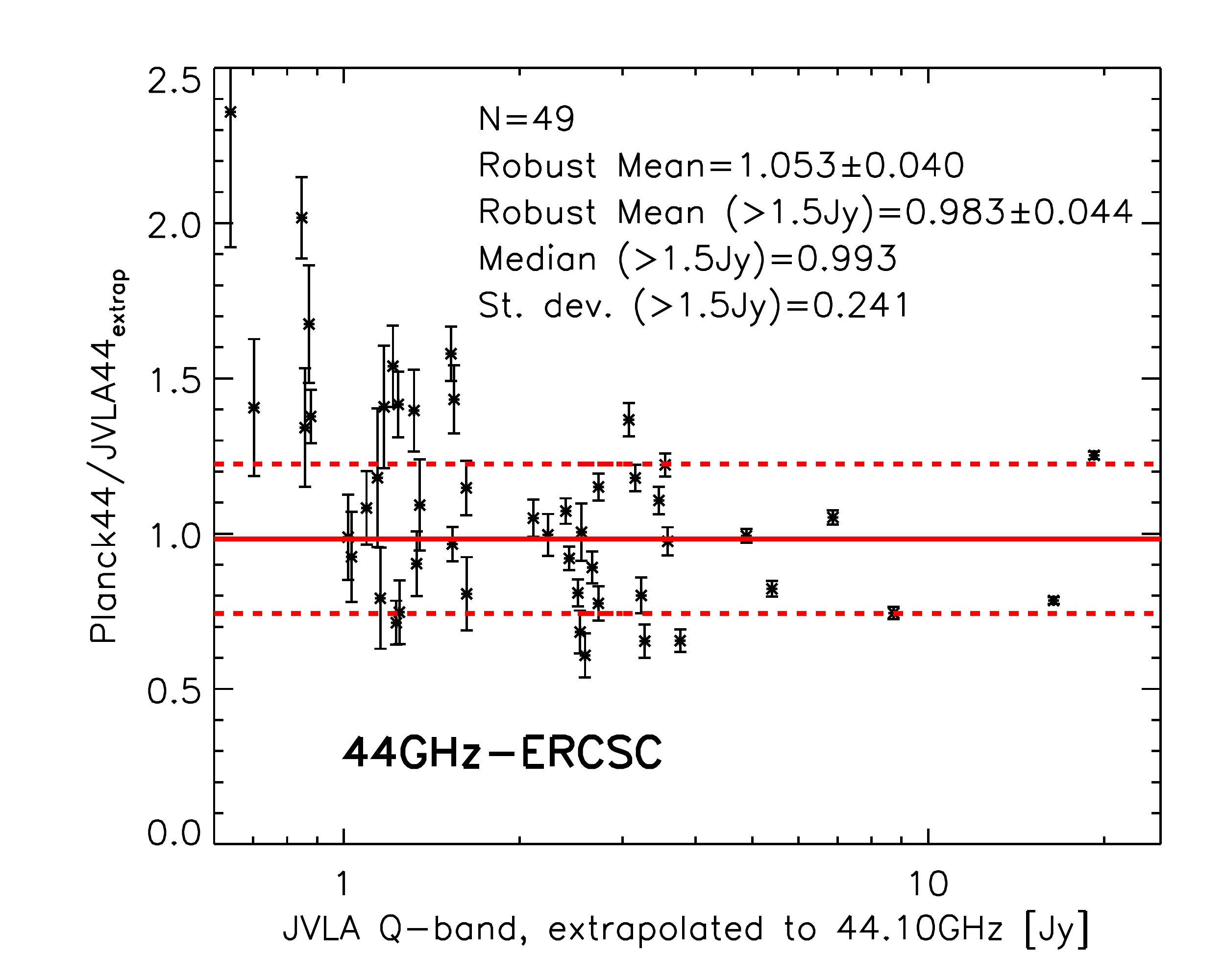}
\caption{The ratio of \Planckns-to-JVLA flux densities vs. JVLA flux densities at 30 (left-hand panel) and 44\,GHz (right-hand panel). The \Planck flux densities used are from the FLUX column of the ERCSC. The JVLA flux densities are extrapolated to the corresponding \Planck central frequencies using the 33-43\,GHz spectral index measured from the JVLA data. Some sources are excluded from this analysis, due to being extended, confused, or highly variable (see text for details). When multiple observations for a given source are available, these are averaged.  Each panel shows the number of sources included, as well as the robust mean, median and standard deviations of the \Planckns-to-JVLA flux density ratios. To mitigate the effects of Eddington bias, we focus on the $>$\,1.5\,Jy sources. The solid red lines show the median ratios for these sources, while the dashed red lines show the +/-1 standard deviation range. }
\label{fig_fluxcomp}
\end{figure*}

Time separations between pairs of VLA/JVLA observations are in the range of 1-450 days. To allow us to look for trends in variability at different time scales, we divide this roughly into three intervals that are big enough to include a statistically significant number of sources. These are: "Interval 1" which covers 0-60 days with a median of 12 days, "Interval 2" which covers 60-260\,days with a median of 168 days, and "Interval 3" which covers 290-430\,days with a median of 359 days. The number of sources included in each interval is $\sim$\,20. Variability is computed for each pair of observations, which are sorted into the appropriate interval depending on their time separation. The vast majority of sources have only 2 separate observations, if any.   In the rare cases where a source has two pairs of observations in the same bin, the two values are averaged. Figure\,\ref{fig_variability} shows the distributions of the variability indices for the different time intervals along with their median values. Note that since the availability and level of multiple epoch observations varies strongly from source to source, this is only representative of the entire sample under the assumption that sources with more repeated observation are not biased toward being more or less variable than sources with fewer repeat observations. This is generally true, with one exception. The most frequently observed sources are the two primary flux calibrators: 3C48 (J0137+3309) and 3C289 (J1331+3030). These sources are used as primary JVLA calibrators precisely because of their very low variability. To remove this bias, we also show the median variabilities with these two sources removed. As expected, the median variabilities are now larger. 

A total of 68 of our sources (76\% of the sample) have been detected by WMAP based on the 7-year data release. This allows us to sample a fourth, longest term interval, "Interval 4", which includes time separation of  1\,--\,6\,years. We make use of the yearly averages for our sources as found in the WMAP 7 year Ka (30\,GHz) data. The time separation between any pair of these varies between 1 and 6 years with a median value of $\sim$\,2.5 years. For the WMAP variability analysis, we apply Equation\,1 with N\,=\,7. 

Figure\,\ref{fig_variability} shows that the median variability of sources increases with timescale from $\sim$\,3\,\% within a few weeks to $\sim$\,14\,\% from year-to-year, with the largest variability ($\sim$\,16\%)  seen in the 1-7 year WMAP data. The fact that there is not a significant difference between our 1-year timescale variability and the 1-7 year WMAP variability estimate, is not surprising given that the median separation between the WMAP measurements is $\sim$\,2.5 years (see above), which is not a large enough separation for us to detect a significant change in the variability.  

We also repeated the analysis for the C-band (5\,GHz) data to test how the variability of our sample changes with frequency as well as timescale. The results are summarized in Table\,\ref{table_var}. As expected, we found that the C-band variability is lower than at Ka, especially on the 1 year timescale where it is $\sim$\,3\,\% at 5\,GHz and 14\,\% at 33\,GHz.  

The result that variability increases with frequency and timescale is consistent with the results found for the PACO bright sample \citep{PACO-proj}, although they only consider time intervals of 90, 180, and 270\,days, whereas we also consider both shorter and longer timescales. They find that the 33\,GHz median variability of their sources increases from 6.7 to 10.6\% between 90 and 270\,days, although dipping slightly to 6.3\% at 180\,days. The 90-270 day range is essentially our "Interval 2" in Figure\,\ref{fig_variability} where we find a median variability of 5.3\% (after excluding the two calibrators).  This is lower than the PACO-bright result, although, given the spread in Figure\,\ref{fig_variability}, the two are marginally consistent. It is somewhat unexpected that PACO should find larger variabilities than we do, given that our sample includes more flat spectrum sources than the PACO-bright sample (see Table\,\ref{table_classes}). The most likely explanation is that the variability index of a given source within a given time interval is based on the spread in flux densities around the mean flux density ($\langle S\rangle$ in Equation\,1) for that particular time interval, whereas \citet{PACO-proj} take the $\langle S\rangle$ to be the mean flux density based on all available observations of a given source. Given that variability increases with timescale, the quoted PACO variabilities are therefore likely to be biased high relative to ours.  We choose to do our analysis in the manner described above, since: 1) the uneven time coverage of our data does not allow us to adopt equivalent "mean flux densities" for all sources; and 2) given the result that variability increases with time, adopting the mean flux density for a larger timescale than considered biases the estimated variability indices for a given interval.  We refer the reader to Chen et al. (2012) for a much more detailed analysis of of the variability of the ERCSC sources based on WMAP data, including a discussion on the limitations of the commonly used variability index formula which we adopt here.

Lastly, Figure\,\ref{fig_obsvar} shows the distribution of time lags between ERCSC observations and each VLA/JVLA observation for the sources in our sample. The two most prominent peaks correspond to approximately six months and one year after the \Planck observations included in the ERCSC. These arise because a typical source is revisited by \Planck every six months. Our earlier VLA observations correspond to the time of \Planck's observations included in the ERCSC. Hence our later JVLA observations, done close in time to \Planck's observations of the same source, occur typically 6 months and then a year after the ERCSC observations. Given the above analysis, we expect the observed variability in comparing the ERCSC and VLA/JVLA observations to be $\sim$\,6-14\%, as shown in Figure\,\ref{fig_variability} and Table\,6. 

\subsection{JVLA-ERCSC Flux Density Comparison}
\label{sec:fluxcomp}

Here we compare the \Planck ERCSC and JVLA flux density scales at 30\,GHz and 44\,GHz. Although our observations were scheduled to be close in time to \Planck observations of the same sources, the comparison here is done against the public ERCSC and our survey extends about a year past the end of the period included in the ERCSC (see Figure\,\ref{fig_obsvar}). In addition, the ERCSC itself includes data averaged over 1.6 sky surveys. Since we lack time ordered data from \Planck for the time of the JVLA follow-up, and given the variability of our sources (see Section\,\ref{sec:variability}), individual source flux comparison is not meaningful. However, a {\it statistical} comparison can still be done. To do so, we extrapolate the Ka (K) ground-based flux densities to the \Planck 30\,GHz and 44\,GHz bands (centered at 28.46\,GHz and 44.1\,GHz respectively). These extrapolations are done by assuming a power law between the Ka (or K) to Q bands with a spectral index as measured for each VLA/JVLA observation of a given source. We also apply color corrections to the \Planck ERCSC flux densities. The multiplicative color corrections from the ERCSC Explanatory Supplement\footnote{http://irsa.ipac.caltech.edu/data/Planck/explanatory\_supplement.pdf} are also based on the Ka-Q spectral index, but as measured from the ERCSC data, whenever both 30 and 44\,GHz matches are found, or from the VLA/JVLA data otherwise. 

As stated in Section\,\ref{sec:match}, 76 of our sources have ERCSC 30\,GHz matches and 59 have 44\,GHz matches.  For the flux density comparison, we exclude M87 which is extended to the JVLA, as well as the two J1310+3220 sources which are blended in the ERCSC. Next, we exclude 8 sources whose WMAP variability index is $>$\,30\,\%, as well as J0359+5057, which is very close to the Galactic plane ($b$\,\,=\,-1.6$^{\circ}$) making its photometry less reliable. All other sources in the sample are sufficiently far from the Galactic plane (see Figure\,\ref{fig_map}) for Galactic emission to be negligible. Finally, we look for remaining outliers, i.e. sources whose ERCSC and JVLA flux densities are more than a factor of 2 discrepant.  There are two such cases: J0418+3801 and J1751+0939. A literature search revealed that J0418+3801 (3C111) is known to be a double lobed radio galaxy \citep{leahy97}. Our measurements are consistent with earlier VLA measurements of the core flux, but \Plancks much larger beam will respond to the extended structure resolved out in our interferometric observations.  As for J1751+0939, our JVLA flux density measurements are below earlier data from both the VLA and WMAP. The source is variable, however, with a WMAP variability index just below our 30\% cutoff mentioned above; therefore, it is possible that our observations were made at a time of particularly weak radio emission. Since our estimates at the four frequency bands are all consistent among themselves, it is less likely that this discrepancy is due to a mistake. We exclude J0418+3801 and J1751+0939 from the flux density comparison. This leaves us with a clean sample for the flux density comparison including 63 sources at 30\,GHz and 49 sources at 44\,GHz. 

Figure\,\ref{fig_fluxcomp} shows the 30 and 44\,GHz JVLA-to-ERCSC flux density comparisons. Whenever applicable, we use the average flux densities based on all our VLA/JVLA observations of the same source. We next want to quantify the level of agreement between the JVLA and \Planck ERCSC flux density scales. In order to minimize the possible effects of any remaining outliers, we compute the robust mean of each ratio, where 3\,$\sigma$ outliers are excluded. At both 30 and 44\,GHz, this removes just 2 sources. At 30\,GHz, the robust mean of the \Planckns-to-JVLA ratio is 1.006\,$\pm$\,0.026, while at 44\,GHz it is 1.053\,$\pm$\,0.040. In both cases, these values are clearly affected by Eddington bias \citep{eddington1913} at the lower flux densities. To minimize this bias, we also compute the robust mean and median values for the sources where the extrapolated JVLA flux density is $>$\,1.5\,Jy. With this flux density cut, no sources were excluded in the calculation of the robust mean. We find the robust mean(median) at 30\,GHz is now 1.008(0.977) and at 44\,GHz it is 0.983(0.993).  

The above analysis suggests that the \Planck ERCSC and (J)VLA flux densities are consistent with each other to within $\sim$\,2-3\% at both 30\,GHz and 44\,GHz.  This level of statistical agreement does not apply to individual sources, whose flux density uncertainties can be much larger. The best example is the primary  (J)VLA flux density calibrator 3C286 (J1331+3030), a well studied source with very low intrinsic variability.  Compared to expectations from the JVLA, 3C286 has a nearly 20\% lower 30\,GHz flux density in the ERCSC\footnote{Although it is in agreement with expectations if the ERCSC's {\sc gauflux} value is used.}. Only a 30\% accuracy in the photometry is required for inclusion in the ERCSC, whose emphasis is on the reliability of the sources. 

Figure\,\ref{fig_fluxcomp} shows significant scatter, which is clearly larger than the nominal ERCSC flux density errors, which in turn are larger than the JVLA errors. Given the median variabilities on a $\sim$\,6 months to a year timescale (see Figure\,\ref{fig_variability}, we predict the scatter to be in the range 6-14\,\%. This is lower than the measured standard deviation of 19\,\%, which may be due to underestimated variability indices as well as additional flux density uncertainties. The nominal ERCSC flux density uncertainties are $\sim$\,3\%, but including systematics they would have to be $\sim$\,15\% to account for the observed scatter. This is within the discrepancy seen between the different flux density estimates (even within the ERCSC) for the essentially non-variable 3C286.    

As noted in Section\,\ref{sec:introduction}, the accuracy of \Planck's calibration on small angular scales, including point sources, is determined by how well the \Planck beam is known -- in essence, the flux density scales as $\theta^{2}$, where $\theta$ is the FWHM of the beam. Here we use the \Planck flux density values as tabulated in the ERCSC, where the FWHM of the 30\,GHz beam is 32.65\,arcmin, and of the 44\,GHz beam it is 27.00\,arcmin. Work is ongoing to further refine our understanding of the \Planck beams. Our results indicate that the \Planck beams are unlikely to be revised by more than $\sim$\,2\%, if we assume the JVLA flux density scale is completely accurate. If they are revised more significantly over time, this may indicate that it is the JVLA flux density scale that needs to be revised. Future work (Perley et al. 2013 in prep.) based on time ordered \Planck data (which also include the observations presented here) will help further constrain the level of agreement between the two instruments. 
  
\section{Summary and Conclusions \label{sec:summary}}

\begin{enumerate}

\item  We present new 5, 8, 33, and 43\,GHz observations for a sample of 89 bright ($>$\,70\% selected to have $S_{37\,GHz}$\,$>$\,1\,Jy) northern radio galaxies. Our observations were scheduled to be within a few weeks  of the \Planck coverage of these intrinsically variable sources, allowing for a range of studies including SED and variability analysis that shed light on the nature of the \Planck detected radio sources. This study is complementary to similar work done in the southern hemisphere by the PACO project \citep[e.g.][]{PACO-proj,PACO-proj-faint} and in the northern hemisphere by the SiMPlE project \citep{procopio11}. Our study has the highest fraction of sources in the ERCSC (85\%), thanks to its selection which favors higher flux density and flatter spectrum sources than PACO or SiMPLE. 

\item From standard color-color diagnostic plots, we find that roughly 1/2 of our sources are flat-spectrum, 1/5 are steep spectrum, 1/5 are down-turning, and the remainder are up-turning, inverted, or self-absorbed. Examination of the 5-857\,GHz SEDs shows that classification on the basis of two spectral indices is over simplistic and varies strongly depending on the frequency examined. 

\item Multiple VLA/JVLA observations for nearly half the sources, along with archival WMAP data, allow us to assess the level of variability in bright radio galaxies on a range of timescales from several weeks to several years. We find that the median variability increases with timescale, from 3\% (about 2 weeks) to 16\% (about 2.5 years).  

\item We compare the flux density scales for \Planckns's two lowest-frequency LFI bands and the JVLA's Ka and Q bands. The agreement between the two instruments is within $\sim$2-3\% at 30\,GHz and 44\,GHz. This is a significant improvement over previously reported results \citep{planck_validation}, thanks to a larger sample, an improved JVLA flux density calibration, and better removal of the intrinsically most variable sources. 

\end{enumerate}

\begin{acknowledgements}
We are grateful to the anonymous referee for a careful reading of the paper and many useful suggestions which have improved the quality and presentation of the paper. We are particularly grateful to the {\sl Planck} Collaboration for providing the public Early Release Compact Source Catalog (ERCSC). A description of the Planck Collaboration and a list of its members, including the technical or scientific activities in which they have been involved, can be found at http://www.sciops.esa.int/index.php?project
=planck\&page=Planck\_Collaboration. This paper makes use of observations obtained at the Expanded Very Large Array (JVLA) which is an instrument of the National Radio Astronomy Observatory (NRAO). The NRAO is a facility of the National Science Foundation operated under cooperative agreement by Associated Universities, Inc. We are especially grateful to Leon Tavares for providing source lists of sources, drawn from the Complete Northern 1\,Jy sample, to be observed by {\sl Planck} in a given week. We are also grateful to Charles Lawrence and Joaquin Gonzales-Nuevo for helpful discussions and reviews of the manuscript before submission. 
This work was funded through a subcontract to Tufts University, and another to Haverford College from NASA/JPL in support of {\sl Planck} related activities.  MLC acknowledges financial support from the Spanish MINECO projects AYA2010-21766-C03-01 and CSD2010-00064. The development of Planck has been supported by: ESA; CNES and CNRS/INSU-IN2P3-INP (France); ASI, CNR, and INAF (Italy); NASA and DoE (USA); STFC and UKSA (UK); CSIC, MICINN and JA (Spain); Tekes, AoF and CSC (Finland); DLR and MPG (Germany); CSA (Canada); DTU Space (Denmark); SER/SSO (Switzerland); RCN (Norway); SFI (Ireland); FCT/MCTES (Portugal); and The development of Planck has been supported by: ESA; CNES and CNRS/INSU-IN2P3-INP (France); ASI, CNR, and INAF (Italy); NASA and DoE (USA); STFC and UKSA (UK); CSIC, MICINN and JA (Spain); Tekes, AoF and CSC (Finland); DLR and MPG (Germany); CSA (Canada); DTU Space (Denmark); SER/SSO (Switzerland); RCN (Norway); SFI (Ireland); FCT/MCTES (Portugal); and PRACE (EU).

\end{acknowledgements}

\bibliographystyle{aa}
\bibliography{nkurinsky}


\clearpage
\onecolumn


\begin{table*}[!h]
\begin{center}
\caption{Table of Scheduling Blocks with bands observed,  dates, and array configuration.  }
\label{sb_table}
\begin{tabular}{c c c c}
\hline
\hline
Scheduling Block & Bands Observed & Date & Configuration \\
\hline
VLA 1 & C, X, K, Q&	Aug 27, 2009 & D  \\
VLA 2 & C, X, K, Q&	Oct 22, 2009 & D  \\
VLA 3 & C,X,K,Q &          Nov 3, 2009 & D  \\
VLA 4 & C, X, K, Q&	Nov 15, 2009 & D  \\
VLA 5 & C, X, K, Q&	Nov 18, 2009 & D  \\
VLA 6 & C, X, K, Q&	Dec 11, 2009 & D  \\
VLA 7 & C, X, K, Q&	Jan 03, 2010 & D  \\
JVLA 1 & C, X, Ka, Q &	Jul 2, 2010 & D  \\
JVLA 2 & C, X, Ka, Q & 	Jul 4, 2010 & D  \\
JVLA 3 & C, X, Ka, Q &	Jul 9, 2010 & D  \\
JVLA 4 & C, X, Ka, Q &	Jul 23, 2010 & D \\
JVLA 5 & C, X, Ka, Q&	Aug 3, 2010 & D \\
JVLA 6 & C, X, Ka, Q &	Sep 7, 2010 & D \\
JVLA 7 & C, X, Ka, Q &	Sep 15, 2010 &	 D \\
JVLA 8 & C, X, Ka, Q &	Sep 20, 2010 &	 DnC \\
JVLA 9 & C, X, Ka, Q &	Sep 24, 2010 &	 DnC \\
JVLA 10 & C, X, Ka, Q &	Oct 18, 2010 & C \\
JVLA 11 & C, X, Ka, Q &	Oct 20, 2010 & C \\
JVLA 12 & C, X, Ka, Q &	Nov 8, 2010 & C \\
JVLA 13 & C, X, Ka, Q &	Nov 9, 2010 & C \\
JVLA 14 &  C, X, Ka, Q &	Nov 20, 2010 &	 C \\
JVLA 15 &  C, X, Ka, Q &	Nov 27, 2010 & C \\
JVLA 16 &  C, X, Ka, Q &	Nov 30, 2010 &	 C \\
\hline
\end{tabular}
\end{center}
\end{table*}


\begin{table*}[!h]
\begin{center}
\caption{The flux densities of the primary JVLA calibrators. The listed correction values ($C_C$, $C_X$, $C_{K}$, and $C_Q$) given have been applied to the previously published VLA data \citep{planck_validation} and represent the ratio of the newer to the older flux density standards (see text for details). }
\label{cal_table}
\begin{tabular}{c | c c | c c c c | c c c c}
\hline
\hline
Cailbrator & RA & Dec & $S_{C}$ (mJy) & $S_X$ (mJy) & $S_{Ka}$ (mJy) & $S_Q$ (mJy) & $C_{C}$ & $C_{X}$ & $C_{K}$ & $C_{Q}$\\
\hline
3C 048 & 01:37:41.2994 & +33.09.35.1330 & 5437 & 3281 & 859 & 623 & 1.020 & 1.041 & 1.134 & 1.268\\
3C 286 & 13:31:08.2880 & +30.30.32.9589 & 7394 & 5207 & 1893 & 1539 & 1.000 & 1.000 & 1.022 & 1.064\\
J0555+3948 & 05:55:30.8056 & +39.48.49.1650 & 5339 & 4836 & 2681 & 2639 & - & - & - & -\\
\hline
\end{tabular}
\end{center}
\end{table*}

\begin{table*}[t]
\caption{VLA/JVLA source coordinates and flux densities. The errors include statistical error as well as calibration uncertainty.  } 
\label{table_reddata}
\begin{center}
\begin{tabular}{crrccccc}
\hline\hline
Source   & RA & Dec & Dateobs & $S_C$ & $S_X$ & $S_{Ka}$ & $S_Q$  \\
 & J2000 & J2000 & & [mJy] & [mJy] & [mJy] & [mJy] \\
\hline
J0006-0623   &   1.5578871  &  -6.3931487  & 20100702  & 2415.0 $\pm$ 9.3  & 2301.6 $\pm$ 14.9  & 1653.4 $\pm$ 8.2  & 1369.1 $\pm$ 10.0 \\
J0019+7327   &   4.9407767  &  73.4583381  & 20090827  & 1191.9 $\pm$ 0.7  & 1224.7 $\pm$ 1.4  & *1051.2 $\pm$ 4.0  & 901.9 $\pm$ 6.2 \\
J0050-0929   &  12.6721558  &  -9.4847806  & 20100723  & 254.1 $\pm$ 1.0  & 209.9 $\pm$ 1.4  & 123.7 $\pm$ 0.6  & 115.2 $\pm$ 0.9 \\
J0108+0135   &  17.1615462  &   1.5834214  & 20100723  & 3558.8 $\pm$ 13.6  & 3909.3 $\pm$ 25.4  & 2736.1 $\pm$ 13.6  & 2295.3 $\pm$ 16.8 \\
J0125-0005   &  21.3701825  &  -0.0988699  & 20100803  & 1263.6 $\pm$ 4.8  & -  & 787.1 $\pm$ 3.9  & 649.7 $\pm$ 4.8 \\
J0137+3309   &  24.4208333  &  33.1597222  & 20090827  & 5537.1 $\pm$ 0.1  & 3267.0 $\pm$ 0.1  & *1234.9 $\pm$ 0.1  & 635.4 $\pm$ 0.1 \\
J0137+3309   &  24.4220808  &  33.1597592  & 20091115  & 5543.7 $\pm$ 0.4  & 3280.7 $\pm$ 0.3  & *1266.6 $\pm$ 5.7  & 667.1 $\pm$ 0.5 \\
J0137+3309   &  24.4220808  &  33.1597592  & 20100702  & 5440.2 $\pm$ 20.8  & 3278.5 $\pm$ 21.3  & 858.1 $\pm$ 4.3  & 670.7 $\pm$ 4.9 \\
J0137+3309   &  24.4220808  &  33.1597592  & 20100723  & 5438.7 $\pm$ 20.8  & -  & 858.9 $\pm$ 4.3  & 669.5 $\pm$ 4.9 \\
J0137+3309   &  24.4220808  &  33.1597592  & 20100803  & 5434.5 $\pm$ 20.8  & -  & 858.1 $\pm$ 4.3  & 669.8 $\pm$ 4.9 \\
J0137+3309   &  24.4220808  &  33.1597592  & 20100907  & 5428.6 $\pm$ 20.8  & 3279.9 $\pm$ 21.3  & 858.4 $\pm$ 4.3  & 670.6 $\pm$ 4.9 \\
J0137+3309   &  24.4220808  &  33.1597592  & 20100920  & 5440.3 $\pm$ 21.0  & -  & 858.8 $\pm$ 4.3  & 666.3 $\pm$ 4.9 \\
J0137+3309   &  24.4220808  &  33.1597592  & 20100924  & 5431.0 $\pm$ 21.2  & -  & 858.5 $\pm$ 4.3  & 684.7 $\pm$ 5.0 \\
J0137+3309   &  24.4220808  &  33.1597592  & 20101020  & 5434.2 $\pm$ 21.2  & -  & 860.7 $\pm$ 4.3  & 672.6 $\pm$ 5.0 \\
J0137+3309   &  24.4220808  &  33.1597592  & 20101109  & 5392.7 $\pm$ 24.5  & -  & 860.5 $\pm$ 4.5  & 672.7 $\pm$ 5.0 \\
J0137+3309   &  24.4220808  &  33.1597592  & 20101120  & 5403.8 $\pm$ 21.9  & -  & 858.9 $\pm$ 4.3  & 674.7 $\pm$ 5.0 \\
J0137+3309   &  24.4220808  &  33.1597592  & 20101127  & 5385.6 $\pm$ 23.5  & 3316.1 $\pm$ 23.0  & 861.8 $\pm$ 4.4  & 674.0 $\pm$ 5.0 \\
J0217+7349   &  34.3783892  &  73.8257283  & 20090827  & 4253.2 $\pm$ 1.0  & 4255.6 $\pm$ 4.2  & *2974.5 $\pm$ 14.2  & 2169.1 $\pm$ 17.8 \\
J0217+7349   &  34.3783892  &  73.8257283  & 20100920  & 3874.4 $\pm$ 14.9  & -  & 2244.3 $\pm$ 11.1  & 1922.5 $\pm$ 14.1 \\
J0217+7349   &  34.3783892  &  73.8257283  & 20100924  & -  & -  & 2284.8 $\pm$ 11.3  & 2284.6 $\pm$ 16.7 \\
J0228+6721   &  37.2085479  &  67.3508415  & 20100920  & 1108.8 $\pm$ 4.3  & -  & 903.6 $\pm$ 4.5  & 861.8 $\pm$ 6.3 \\
J0319+4130   &  49.9506671  &  41.5116953  & 20100907  & 14555.1 $\pm$ 56.1  & 21103.0 $\pm$ 137.0  & 18627.6 $\pm$ 92.4  & 16539.0 $\pm$ 121.0 \\
J0336+3218   &  54.1254483  &  32.3081507  & 20100907  & -  & 2095.6 $\pm$ 13.6  & 2843.4 $\pm$ 14.1  & 2571.6 $\pm$ 18.8 \\
J0359+5057   &  59.8739471  &  50.9639337  & 20100920  & 8834.3 $\pm$ 33.8  & -  & 8246.9 $\pm$ 40.9  & 7162.9 $\pm$ 52.4 \\
J0418+3801   &  64.5886542  &  38.0266111  & 20100920  & 2827.4 $\pm$ 19.6  & -  & 2233.3 $\pm$ 11.1  & 2077.3 $\pm$ 15.2 \\
J0423-0120   &  65.8125000  &  -1.3425000  & 20091103  & -  & -  & *8505.0 $\pm$ 0.1  & 9245.7 $\pm$ 0.1 \\
J0423-0120   &  65.8157917  &  -1.3426111  & 20091103  & 3966.9 $\pm$ 0.5  & 4398.9 $\pm$ 0.5  & *8585.7 $\pm$ 0.6  & 9246.5 $\pm$ 0.6 \\
J0423-0120   &  65.8158363  &  -1.3425181  & 20100907  & -  & 6200.0 $\pm$ 40.2  & 5651.5 $\pm$ 28.0  & 5183.3 $\pm$ 37.9 \\
J0433+0521   &  68.2962312  &   5.3543387  & 20100907  & 3225.1 $\pm$ 12.4  & 2765.8 $\pm$ 18.0  & 1539.1 $\pm$ 7.7  & 1333.7 $\pm$ 9.8 \\
J0449+1121   &  72.2819629  &  11.3579435  & 20100907  & 1175.2 $\pm$ 4.5  & 1323.4 $\pm$ 8.6  & 1254.8 $\pm$ 6.2  & 1126.1 $\pm$ 8.2 \\
J0449+1121   &  72.2819629  &  11.3579435  & 20100920  & 1161.5 $\pm$ 4.5  & -  & 1248.4 $\pm$ 6.2  & 1173.3 $\pm$ 8.6 \\
J0501-0159   &  75.3375000  &  -1.9872934  & 20100907  & 1037.5 $\pm$ 4.0  & 1158.7 $\pm$ 7.5  & 1497.5 $\pm$ 7.4  & 1382.2 $\pm$ 10.1 \\
J0510+1800   &  77.5098713  &  18.0115505  & 2010920  & -  & -  & 565.3 $\pm$ 2.8  & 584.6 $\pm$ 4.3 \\
J0530+1331   &  82.7350696  &  13.5319860  & 20100920  & 2191.4 $\pm$ 8.4  & -  & 1119.8 $\pm$ 5.6  & 1098.3 $\pm$ 8.0 \\
J0530+1331   &  82.7350696  &  13.5319860  & 20100924  & 2247.6 $\pm$ 8.6  & -  & 1140.4 $\pm$ 5.7  & 1239.2 $\pm$ 9.1 \\
J0541-0541   &  85.4086808  &  -5.6970634  & 20101130  & 2035.6 $\pm$ 1.0  & 1046.7 $\pm$ 1.0  & 582.0 $\pm$ 0.2  & - \\
J0555+3948   &  88.8783567  &  39.8136569  & 20100301  & 5338.6  & 4835.9  & 2681.1  & 2639.4 \\
J0555+3948   &  88.8783567  &  39.8136569  & 20100924  & 5199.3 $\pm$ 19.9  & -  & 2453.9 $\pm$ 12.2  & 2461.5 $\pm$ 18.0 \\
J0555+3948   &  88.8783567  &  39.8136569  & 20101018  & 5337.5 $\pm$ 0.3  & -  & 2682.4 $\pm$ 0.7  & 2622.3 $\pm$ 1.7 \\
J0555+3948   &  88.8783567  &  39.8136569  & 20101130  & 5069.1 $\pm$ 8.8  & 4738.7 $\pm$ 4.6  & 2681.1 $\pm$ 0.4  & 2635.9 $\pm$ 0.6 \\
J0607-0834   &  91.9987467  &  -8.5805495  & 20100924  & -  & -  & 2237.8 $\pm$ 11.1  & 2235.1 $\pm$ 16.4 \\
J0646+4451   & 101.6334417  &  44.8546084  & 20100924  & -  & -  & 2532.5 $\pm$ 12.6  & 2537.1 $\pm$ 18.6 \\
J0721+7120   & 110.4727021  &  71.3434343  & 20101018  & 1738.2 $\pm$ 2.6  & -  & 4024.2 $\pm$ 1.3  & 4251.5 $\pm$ 2.2 \\
J0725-0054   & 111.4610000  &  -0.9156944  & 20091103  & -  & -  & *2571.0 $\pm$ 0.1  & 2365.0 $\pm$ 0.1 \\
J0725-0054   & 111.4610000  &  -0.9157068  & 20101018  & 3451.3 $\pm$ 1.7  & 4073.0 $\pm$ 0.5  & 4045.8 $\pm$ 1.2  & 4093.5 $\pm$ 0.7 \\
J0738+1742   & 114.5308071  &  17.7052773  & 20101018  & 797.5 $\pm$ 1.8  & 784.1 $\pm$ 0.7  & 580.8 $\pm$ 0.3  & 591.2 $\pm$ 0.8 \\
J0739+0137   & 114.8251408  &   1.6179494  & 20091022  & 1143.0 $\pm$ 10.0  & 1260.0 $\pm$ 1.3  & *1546.7 $\pm$ 4.1  & 1652.8 $\pm$ 3.2 \\
J0739+0137   & 114.8251412  &   1.6179494  & 20101018  & 833.7 $\pm$ 1.1  & 866.2 $\pm$ 0.4  & 993.6 $\pm$ 0.5  & 1042.9 $\pm$ 0.7 \\
J0745-0044   & 116.4750000  &  -0.7380556  & 20091103  & 1970.0 $\pm$ 0.1  & 1964.0 $\pm$ 0.1  & *960.0 $\pm$ 0.1  & 605.0 $\pm$ 0.1 \\
J0750+1231   & 117.7166667  &  12.5177778  & 20091103  & 3685.0 $\pm$ 0.1  & 4486.0 $\pm$ 0.1  & *4400.0 $\pm$ 0.1  & 3940.0 $\pm$ 0.1 \\
J0750+1231   & 117.7168571  &  12.5180078  & 20101018  & 3750.8 $\pm$ 0.5  & -  & 3079.0 $\pm$ 0.9  & 3002.3 $\pm$ 1.4 \\
J0757+0956   & 119.2776667  &   9.9430278  & 20091022  & 1143.0 $\pm$ 6.0  & 1530.0 $\pm$ 1.2  & *1903.2 $\pm$ 4.1  & 1964.7 $\pm$ 5.2 \\
J0757+0956   & 119.2776787  &   9.9430145  & 20101018  & 1026.5 $\pm$ 0.3  & 1185.0 $\pm$ 1.2  & 1250.4 $\pm$ 0.4  & 1276.4 $\pm$ 1.3 \\
J0808+4950   & 122.1652758  &  49.8434806  & 20091022  & 464.6 $\pm$ 0.9  & 454.0 $\pm$ 0.4  & *488.0 $\pm$ 1.3  & 4220.6 $\pm$ 2.1 \\
J0825+0309   & 126.4595833  &   3.1566667  & 20091022  & 776.9 $\pm$ 0.7  & 737.4 $\pm$ 0.6  & *682.1 $\pm$ 3.5  & 689.7 $\pm$ 2.3 \\
J0825+0309   & 126.4597433  &   3.1568111  & 20101108  & 559.6 $\pm$ 5.4  & -  & 925.8 $\pm$ 2.4  & 1005.9 $\pm$ 1.1 \\
J0830+2410   & 127.7170000  &  24.1833333  & 20091022  & 1269.0 $\pm$ 7.0  & 1246.0 $\pm$ 1.0  & *1072.7 $\pm$ 5.1  & 877.4 $\pm$ 4.2 \\
J0841+7053   & 130.3515417  &  70.8950556  & 20091022  & 1762.0 $\pm$ 9.0  & 1484.5 $\pm$ 0.8  & *2267.0 $\pm$ 4.7  & 2900.0 $\pm$ 9.4 \\
J0841+7053   & 130.3515221  &  70.8950481  & 20101018  & 1872.7 $\pm$ 0.6  & -  & 2905.8 $\pm$ 0.9  & 2559.4 $\pm$ 1.5 \\
J0854+2006   & 133.7036454  &  20.1085114  & 20101108  & 3531.4 $\pm$ 33.9  & -  & 5997.9 $\pm$ 15.4  & 6803.7 $\pm$ 7.1 \\
J0909+4253   & 137.3895833  &  42.8961389  & 20091118  & 1772.6 $\pm$ 0.6  & 1371.1 $\pm$ 1.1  & *973.5 $\pm$ 1.6  & 817.1 $\pm$ 2.2 \\
J0920+4441   & 140.2435771  &  44.6983292  & 20091118  & 1124.5 $\pm$ 0.4  & 1518.3 $\pm$ 1.0  & *2475.3 $\pm$ 0.6  & 2393.0 $\pm$ 1.0 \\
J0920+4441   & 140.2435771  &  44.6983292  & 20091118  & 1130.2 $\pm$ 0.5  & 1530.5 $\pm$ 0.6  & *2478.5 $\pm$ 1.7  & 2495.2 $\pm$ 5.5 \\
J0920+4441   & 140.2435771  &  44.6983292  & 20101108  & 1138.9 $\pm$ 10.9  & -  & 2758.0 $\pm$ 7.1  & 3251.8 $\pm$ 3.5 \\
J0927+3902   & 141.7625579  &  39.0391255  & 20091118  & 11092.0 $\pm$ 3.5  & 11846.0 $\pm$ 6.7  & *9388.5 $\pm$ 40.9  & 7460.0 $\pm$ 31.9 \\
J0927+3902   & 141.7625579  &  39.0391255  & 20101108  & 10441.0 $\pm$ 100.8  & -  & 9309.7 $\pm$ 23.9  & 9996.1 $\pm$ 10.5 \\
J0948+4039   & 147.2305754  &  40.6623853  & 20091118  & 1656.3 $\pm$ 0.8  & 1563.0 $\pm$ 1.3  & *1430.0 $\pm$ 20.0  & 1130.0 $\pm$ 10.0 \\
J0948+4039   & 147.2305754  &  40.6623853  & 20101108  & 1772.0 $\pm$ 19.5  & -  & 1151.7 $\pm$ 3.0  & 1293.9 $\pm$ 1.5 \\
\hline
\end{tabular}
\end{center}
\end{table*}

\setcounter{table}{2}

\begin{table*}[t]
\caption{ \it{Continued}} 
\begin{center}
\begin{tabular}{cccccccc}
\hline\hline
Source   & RA & Dec & Dateobs & $S_C$ & $S_X$ & $S_{Ka}$ & $S_Q$  \\
 & J2000 & J2000 & & [mJy] & [mJy] & [mJy] & [mJy] \\
\hline
J0956+2515   & 149.2078142  &  25.2544583  & 20101108  & 667.0 $\pm$ 6.3  & -  & 1230.9 $\pm$ 3.2  & 1316.2 $\pm$ 1.5 \\
J0956+2515   & 149.2078142  &  25.2544583  & 20101130  & 642.6 $\pm$ 1.2  & 699.9 $\pm$ 0.8  & 1468.7 $\pm$ 0.2  & 1757.4 $\pm$ 0.6 \\
J1038+0512   & 159.6949167  &   5.2080806  & 20100103  & 1233.0 $\pm$ 29.0  & 1920.0 $\pm$ 1.3  & *1513.0 $\pm$ 3.5  & 1164.2 $\pm$ 3.6 \\
J1043+2408   & 160.7876492  &  24.1431693  & 20101130  & 644.5 $\pm$ 0.2  & 742.9 $\pm$ 1.0  & 1149.2 $\pm$ 0.2  & 1299.0 $\pm$ 0.5 \\
J1058+0234   & 164.6208333  &   2.5830556  & 20091103  & -  & -  & *5985.0 $\pm$ 0.1  & 6344.0 $\pm$ 0.1 \\
J1058+0133   & 164.6233542  &   1.5663333  & 20100103  & 3193.0 $\pm$ 7.0  & 4186.0 $\pm$ 3.0  & *5431.9 $\pm$ 13.3  & 5404.0 $\pm$ 24.5 \\
J1130+3815   & 172.7220000  &  38.2551389  & 20100103  & 1334.0 $\pm$ 2.0  & 1356.0 $\pm$ 1.0  & *1041.0 $\pm$ 1.5  & 723.7 $\pm$ 3.6 \\
J1130+3815   & 172.7220108  &  38.2551520  & 20101130  & 1370.6 $\pm$ 2.7  & 1538.7 $\pm$ 1.7  & 1327.8 $\pm$ 0.4  & 1319.5 $\pm$ 0.5 \\
J1153+8043   & 178.3500000  &  80.7311111  & 20091103  & -  & -  & *803.0 $\pm$ 0.1  & 642.0 $\pm$ 0.1 \\
J1153+4931   & 178.3519583  &  49.5191194  & 20100103  & 1167.0 $\pm$ 1.2  & 949.0 $\pm$ 0.9  & *904.1 $\pm$ 0.8  & 849.2 $\pm$ 2.4 \\
J1159+2914   & 179.8791667  &  29.2452778  & 20091103  & 1609.0 $\pm$ 0.1  & 1530.0 $\pm$ 0.1  & *1416.0 $\pm$ 0.1  & 1140.0 $\pm$ 0.1 \\
J1159+2914   & 179.8826250  &  29.2455556  & 20100103  & 1609.0 $\pm$ 1.3  & 1530.0 $\pm$ 3.0  & *1446.6 $\pm$ 4.1  & 1214.3 $\pm$ 9.6 \\
J1221+2813   & 185.3820438  &  28.2329168  & 201074  & 446.6 $\pm$ 4.2  & 419.3 $\pm$ 2.2  & 326.7 $\pm$ 0.8  & 297.5 $\pm$ 0.4 \\
J1222+0413   & 185.5939567  &   4.2210490  & 20100704  & 783.7 $\pm$ 7.4  & 1076.3 $\pm$ 5.7  & 1000.8 $\pm$ 2.6  & 903.6 $\pm$ 1.1 \\
J1222+0413   & 185.5939567  &   4.2210490  & 20100709  & 782.8 $\pm$ 7.4  & 1089.7 $\pm$ 5.7  & 1063.5 $\pm$ 2.7  & 1021.7 $\pm$ 1.1 \\
J1224+2122   & 186.2271250  &  21.3797222  & 20100103  & 1233.0 $\pm$ 2.0  & 1267.0 $\pm$ 1.4  & *1442.5 $\pm$ 4.1  & 1260.0 $\pm$ 8.5 \\
J1229+0203   & 187.2779154  &   2.0523884  & 20100704  & 37172.1 $\pm$ 360.1  & 31724.1 $\pm$ 187.0  & 22899.1 $\pm$ 59.9  & 19226.6 $\pm$ 24.5 \\
J1229+0203   & 187.2779154  &   2.0523884  & 20100709  & 37738.1 $\pm$ 357.1  & 32793.1 $\pm$ 173.6  & 24318.3 $\pm$ 62.4  & 19812.5 $\pm$ 20.3 \\
J1230+1223   & 187.7059308  &  12.3911233  & 20100704  & 53260.1 $\pm$ 548.8  & 24892.1 $\pm$ 163.9  & 1669.4 $\pm$ 14.2  & 1807.8 $\pm$ 7.7 \\
J1230+1223   & 187.7059308  &  12.3911233  & 20100709  & 59530.1 $\pm$ 719.9  & 26430.1 $\pm$ 177.1  & 1913.7 $\pm$ 13.5  & 1884.1 $\pm$ 5.0 \\
J1310+3220   & 197.6194167  &  32.3455000  & 20100103  & 1427.0 $\pm$ 1.7  & 2133.0 $\pm$ 1.4  & *2917.7 $\pm$ 5.1  & 2125.2 $\pm$ 16.0 \\
J1310+3220   & 197.6194325  &  32.3454953  & 20100709  & 1963.4 $\pm$ 18.5  & 2718.1 $\pm$ 14.3  & 2548.0 $\pm$ 6.5  & 2312.3 $\pm$ 2.4 \\
J1327+2210   & 201.7535879  &  22.1806010  & 20100709  & 1856.9 $\pm$ 17.5  & 1746.1 $\pm$ 9.2  & 1232.9 $\pm$ 3.2  & 1077.2 $\pm$ 1.1 \\
J1331+3030   & 202.7845333  &  30.5091553  & 20091118  & 7481.4 $\pm$ 0.6  & 5202.6 $\pm$ 0.5  & *2574.5 $\pm$ 1.6  & 1557.1 $\pm$ 0.5 \\
J1331+3030   & 202.7845333  &  30.5091553  & 20100103  & 7485.0 $\pm$ 0.5  & 5202.3 $\pm$ 0.4  & *2570.4 $\pm$ 0.3  & 1532.5 $\pm$ 0.3 \\
J1331+3030   & 202.7845333  &  30.5091553  & 20100704  & 7398.4 $\pm$ 69.8  & 5200.4 $\pm$ 27.3  & 1890.9 $\pm$ 4.9  & 1530.3 $\pm$ 1.7 \\
J1331+3030   & 202.7845333  &  30.5091553  & 20100709  & 7387.3 $\pm$ 69.7  & 5243.5 $\pm$ 27.5  & 1891.3 $\pm$ 4.9  & 1531.1 $\pm$ 1.6 \\
J1331+3030   & 202.7845333  &  30.5091553  & 20100915  & 7382.4 $\pm$ 69.7  & 5251.5 $\pm$ 27.6  & 1898.6 $\pm$ 4.9  & 1533.8 $\pm$ 1.7 \\
J1331+3030   & 202.7845333  &  30.5091553  & 20101108  & 7251.2 $\pm$ 68.7  & -  & 1892.2 $\pm$ 4.9  & 1532.6 $\pm$ 1.6 \\
J1419+5423   & 214.9441667  &  54.3872222  & 20100103  & 1178.0 $\pm$ 1.6  & 1249.0 $\pm$ 7.0  & *1257.6 $\pm$ 1.6  & 1251.5 $\pm$ 4.3 \\
J1419+5423   & 214.9441558  &  54.3874409  & 20100704  & 1097.6 $\pm$ 10.4  & 1096.9 $\pm$ 5.8  & 963.6 $\pm$ 2.5  & 949.8 $\pm$ 1.1 \\
J1419+5423   & 214.9441558  &  54.3874409  & 20100709  & 1063.2 $\pm$ 10.1  & 1081.2 $\pm$ 5.7  & 909.7 $\pm$ 2.3  & 893.3 $\pm$ 0.9 \\
J1642+6856   & 250.5327021  &  68.9443768  & 20100704  & 2520.8 $\pm$ 23.8  & 2617.4 $\pm$ 13.7  & 1696.2 $\pm$ 4.9  & 1524.8 $\pm$ 2.1 \\
J1642+6856   & 250.5327021  &  68.9443768  & 20100709  & 2482.9 $\pm$ 23.4  & -  & 1735.3 $\pm$ 4.5  & 1567.7 $\pm$ 1.6 \\
J1728+1215   & 262.0293800  &  12.2609683  & 2010915  & -  & 464.8 $\pm$ 2.4  & 451.8 $\pm$ 1.2  & 425.5 $\pm$ 0.5 \\
J1743-0350   & 265.9952337  &  -3.8346158  & 20100915  & 3316.3 $\pm$ 31.3  & 3469.5 $\pm$ 18.2  & 3616.1 $\pm$ 9.3  & 3298.0 $\pm$ 3.5 \\
J1751+0939   & 267.8867442  &   9.6502024  & 20100915  & 1619.8 $\pm$ 15.3  & 1905.7 $\pm$ 10.0  & 2111.2 $\pm$ 5.5  & 2085.1 $\pm$ 2.2 \\
J1800+3848   & 270.1031892  &  38.8085271  & 20100915  & -  & -  & 551.6 $\pm$ 1.4  & 504.8 $\pm$ 0.6 \\
J1800+7828   & 270.1903496  &  78.4677829  & 20101020  & 2497.1 $\pm$ 9.6  & -  & 2522.4 $\pm$ 12.5  & 2515.8 $\pm$ 18.4 \\
J1806+6949   & 271.7111692  &  69.8244746  & 20100704  & 1537.1 $\pm$ 14.7  & 1478.1 $\pm$ 7.8  & 1202.0 $\pm$ 3.5  & 1185.5 $\pm$ 1.9 \\
J1806+6949   & 271.7111692  &  69.8244746  & 20100709  & 1524.5 $\pm$ 14.6  & 1481.1 $\pm$ 7.8  & 1235.2 $\pm$ 3.4  & 1192.5 $\pm$ 1.8 \\
J1806+6949   & 271.7111692  &  69.8244746  & 20101109  & 1383.4 $\pm$ 6.4  & -  & 1241.0 $\pm$ 6.3  & 1296.0 $\pm$ 9.5 \\
J1806+6949   & 271.7111692  &  69.8244746  & 20101127  & 1440.8 $\pm$ 6.3  & 1409.2 $\pm$ 10.3  & 1058.7 $\pm$ 5.3  & 1231.3 $\pm$ 9.1 \\
J1824+5651   & 276.0294517  &  56.8504141  & 20100915  & -  & 1362.2 $\pm$ 7.2  & 1137.2 $\pm$ 3.1  & 1096.9 $\pm$ 1.4 \\
J1829+4844   & 277.3824300  &  48.7461558  & 20100915  & 5042.5 $\pm$ 47.6  & 3716.4 $\pm$ 19.5  & 2498.6 $\pm$ 6.8  & 2405.8 $\pm$ 4.4 \\
J1849+6705   & 282.3169679  &  67.0949111  & 20100723  & 1434.7 $\pm$ 5.5  & 2152.3 $\pm$ 14.0  & 2720.4 $\pm$ 13.5  & 2473.9 $\pm$ 18.1 \\
J1849+6705   & 282.3169679  &  67.0949111  & 20100803  & 1452.0 $\pm$ 5.7  & -  & 2979.9 $\pm$ 14.8  & 2861.2 $\pm$ 20.9 \\
J1849+6705   & 282.3169679  &  67.0949111  & 20100915  & 1395.9 $\pm$ 13.2  & 2050.6 $\pm$ 10.8  & 2463.8 $\pm$ 6.3  & 2212.1 $\pm$ 2.3 \\
J1849+6705   & 282.3169679  &  67.0949111  & 20101020  & -  & -  & 2252.6 $\pm$ 11.2  & 2349.2 $\pm$ 17.2 \\
J1849+6705   & 282.3169679  &  67.0949111  & 20101109  & 1259.9 $\pm$ 5.0  & -  & 2247.8 $\pm$ 11.2  & 2295.2 $\pm$ 16.8 \\
J1927+7358   & 291.9520633  &  73.9671028  & 20101020  & -  & -  & 4446.9 $\pm$ 22.2  & 4851.7 $\pm$ 35.5 \\
J1955+5131   & 298.9280763  &  51.5301517  & 20101020  & -  & -  & 1333.4 $\pm$ 6.6  & 1589.4 $\pm$ 11.6 \\
J1955+5131   & 298.9280763  &  51.5301517  & 20101120  & 1296.5 $\pm$ 5.7  & -  & 1267.8 $\pm$ 6.3  & 1127.7 $\pm$ 8.3 \\
J1955+5131   & 298.9280763  &  51.5301517  & 20101127  & 1326.8 $\pm$ 5.9  & 1387.5 $\pm$ 9.2  & 1141.8 $\pm$ 5.7  & 1015.4 $\pm$ 7.4 \\
J2005+7752   & 301.3791604  &  77.8786799  & 20101020  & 924.7 $\pm$ 3.6  & -  & 689.6 $\pm$ 3.4  & 701.0 $\pm$ 5.2 \\
J2007+4029   & 301.9372704  &  40.4968345  & 20091115  & 2956.5 $\pm$ 3.1  & 3810.3 $\pm$ 1.9  & *3141.2 $\pm$ 14.7  & 2465.4 $\pm$ 20.3 \\
J2007+4029   & 301.9372704  &  40.4968345  & 20101020  & 3534.2 $\pm$ 13.6  & -  & 3506.0 $\pm$ 17.4  & 3949.9 $\pm$ 29.0 \\
J2022+6136   & 305.5278404  &  61.6163346  & 20100702  & 3161.5 $\pm$ 12.1  & 3058.8 $\pm$ 19.9  & 1136.0 $\pm$ 5.6  & 839.6 $\pm$ 6.1 \\
J2022+6136   & 305.5278404  &  61.6163346  & 20100723  & 3227.4 $\pm$ 12.4  & 3089.9 $\pm$ 20.1  & 1134.5 $\pm$ 5.6  & 819.8 $\pm$ 6.0 \\
J2022+6136   & 305.5278404  &  61.6163346  & 20100803  & 3242.4 $\pm$ 12.4  & -  & 1213.7 $\pm$ 6.0  & 1027.5 $\pm$ 7.5 \\
J2053+5427   & 313.4791667  &  54.4597222  & 20091103  & 2850.0 $\pm$ 0.1  & 3311.0 $\pm$ 0.1  & *2460.0 $\pm$ 0.1  & 1360.0 $\pm$ 0.1 \\
J2123+0535   & 320.9354892  &   5.5894703  & 20091115  & 2168.4 $\pm$ 4.0  & 1927.0 $\pm$ 2.0  & *1144.8 $\pm$ 6.5  & 847.5 $\pm$ 9.3 \\
J2123+0535   & 320.9354892  &   5.5894703  & 20101109  & 2564.6 $\pm$ 10.9  & -  & 1421.1 $\pm$ 7.1  & 1550.6 $\pm$ 11.3 \\
J2133+1443   & 323.4057888  &  14.7295753  & 20101109  & 164.5 $\pm$ 0.7  & -  & 133.7 $\pm$ 0.7  & 118.6 $\pm$ 0.9 \\
J2134-0153   & 323.5429583  &  -1.8881111  & 20091103  & -  & -  & *1925.5 $\pm$ 0.1  & 1545.0 $\pm$ 0.1 \\
J2134-0153   & 323.5429583  &  -1.8881111  & 20091115  & 2450.2 $\pm$ 5.4  & 2492.0 $\pm$ 2.7  & *2089.5 $\pm$ 10.1  & 1717.1 $\pm$ 14.6 \\
J2136+0041   & 324.1607762  &   0.6983926  & 20091115  & -  & 8334.4 $\pm$ 35.4  & *4367.0 $\pm$ 27.2  & 2536.1 $\pm$ 26.3 \\
J2136+0041   & 324.1607762  &   0.6983926  & 20101109  & 9412.6 $\pm$ 36.8  & -  & 4968.0 $\pm$ 24.7  & 5028.3 $\pm$ 36.8 \\
\hline
\end{tabular}
\end{center}
\end{table*}

\setcounter{table}{2}

\begin{table*}[t]
\caption{ \it{Continued}} 
\begin{center}
\begin{tabular}{cccccccc}
\hline\hline
Source   & RA & Dec & Dateobs & $S_C$ & $S_X$ & $S_{Ka}^*$ & $S_Q$  \\
 & J2000 & J2000 & & [mJy] & [mJy] & [mJy] & [mJy] \\
\hline
J2139+1423   & 324.7554554  &  14.3933311  & 20091115  & 2986.9 $\pm$ 5.0  & 3202.7 $\pm$ 2.1  & *2080.3 $\pm$ 9.3  & 1294.1 $\pm$ 14.7 \\
J2139+1423   & 324.7554554  &  14.3933311  & 20101120  & 2930.4 $\pm$ 11.9  & -  & 2015.1 $\pm$ 10.0  & 1595.0 $\pm$ 11.7 \\
J2139+1423   & 324.7554554  &  14.3933311  & 20101127  & 2987.5 $\pm$ 11.7  & 3345.4 $\pm$ 21.8  & 1613.0 $\pm$ 8.0  & 1726.4 $\pm$ 12.6 \\
J2148+0657   & 327.0227500  &   6.9607222  & 20091115  & 5356.3 $\pm$ 15.3  & 5559.4 $\pm$ 7.6  & *3852.2 $\pm$ 27.2  & 3172.3 $\pm$ 33.0 \\
J2148+6107   & 327.0668558  &  61.1182883  & 20090827  & 1224.7 $\pm$ 0.6  & 996.1 $\pm$ 1.3  & *601.1 $\pm$ 2.5  & 330.0 $\pm$ 4.2 \\
J2202+4216   & 330.6804167  &  42.2775000  & 20091103  & 3994.0 $\pm$ 0.1  & 3996.0 $\pm$ 0.1  & *3440.0 $\pm$ 0.1  & 3085.0 $\pm$ 0.1 \\
J2203+3145   & 330.8124167  &  31.7606389  & 20091103  & 2165.0 $\pm$ 0.1  & 2495.0 $\pm$ 0.1  & *2576.0 $\pm$ 0.1  & 2841.0 $\pm$ 0.1 \\
J2203+3145   & 330.8123992  &  31.7606305  & 20100702  & 1997.6 $\pm$ 8.3  & 2515.4 $\pm$ 16.3  & 2864.1 $\pm$ 14.2  & 2500.8 $\pm$ 18.3 \\
J2203+1725   & 330.8620417  &  17.4300556  & 20091103  & 987.0 $\pm$ 0.1  & 1055.0 $\pm$ 0.1  & *1035.0 $\pm$ 0.1  & 1073.0 $\pm$ 0.1 \\
J2203+1725   & 330.8620571  &  17.4300688  & 20101120  & 937.2 $\pm$ 4.6  & -  & 1573.3 $\pm$ 7.8  & 1271.6 $\pm$ 9.3 \\
J2203+1725   & 330.8620571  &  17.4300688  & 20101127  & 893.4 $\pm$ 3.6  & 1197.2 $\pm$ 7.8  & 1558.5 $\pm$ 7.7  & 1663.6 $\pm$ 12.2 \\
J2218-0335   & 334.7168333  &  -3.5935833  & 20091115  & 2254.1 $\pm$ 3.9  & 1875.0 $\pm$ 1.6  & *1214.3 $\pm$ 4.8  & 1021.5 $\pm$ 8.2 \\
J2225-0457   & 336.4469167  &  -4.9503889  & 20091115  & 7219.7 $\pm$ 16.0  & 7795.0 $\pm$ 5.0  & *5656.0 $\pm$ 24.0  & 3597.0 $\pm$ 26.0 \\
J2229-0832   & 337.4170417  &  -8.5484722  & 20091115  & 2085.4 $\pm$ 4.5  & 1767.6 $\pm$ 1.2  & *1234.3 $\pm$ 5.4  & 1239.3 $\pm$ 9.3 \\
J2230+6946   & 337.6519571  &  69.7744658  & 20090827  & 743.1 $\pm$ 0.4  & 748.2 $\pm$ 0.9  & *572.7 $\pm$ 3.3  & 405.9 $\pm$ 3.7 \\
J2232+1143   & 338.1517083  &  11.7308056  & 20091103  & 5594.0 $\pm$ 0.1  & 5827.0 $\pm$ 0.1  & *4407.0 $\pm$ 0.1  & 3606.0 $\pm$ 0.1 \\
J2236+2828   & 339.0936250  &  28.4826111  & 20091103  & 1285.0 $\pm$ 0.1  & 1273.0 $\pm$ 0.1  & *1412.0 $\pm$ 0.1  & 1547.0 $\pm$ 0.1 \\
J2236+2828   & 339.0936287  &  28.4826148  & 20100702  & 1595.7 $\pm$ 6.1  & 1434.7 $\pm$ 9.3  & 1595.7 $\pm$ 7.9  & 1541.4 $\pm$ 11.3 \\
J2253+1608   & 343.4906250  &  16.1482222  & 20091103  & 9463.0 $\pm$ 0.1  & 8176.0 $\pm$ 0.1  & *11834.0 $\pm$ 0.1  & 21570.1 $\pm$ 0.1 \\
J2253+1608   & 343.4906162  &  16.1482114  & 20100702  & 12572.6 $\pm$ 48.2  & -  & 28126.0 $\pm$139.5  & 26563.1 $\pm$ 194.2 \\
J2316+1618   & 349.1654112  &  16.3018731  & 201072  & 246.1 $\pm$ 1.0  & 198.1 $\pm$ 1.3  & 139.3 $\pm$ 0.7  & 129.6 $\pm$ 1.0 \\
J2327+0940   & 351.8899192  &   9.6692952  & 20100702  & 1024.9 $\pm$ 3.9  & 1002.9 $\pm$ 6.5  & 990.4 $\pm$ 4.9  & 891.8 $\pm$ 6.5 \\
J2327+0940   & 351.8899192  &   9.6692952  & 20100723  & 1010.1 $\pm$ 3.9  & 1012.2 $\pm$ 6.6  & 1123.2 $\pm$ 5.6  & 1040.7 $\pm$ 7.6 \\
J2327+0940   & 351.8899192  &   9.6692952  & 20100803  & 998.8 $\pm$ 3.8  & -  & 1317.3 $\pm$ 6.5  & 1300.0 $\pm$ 9.5 \\
J2354+4553   & 358.5903346  &  45.8845101  & 20100723  & 1260.0 $\pm$ 4.8  & 1012.9 $\pm$ 6.6  & 492.5 $\pm$ 2.4  & 399.5 $\pm$ 2.9 \\
J2354+4553   & 358.5903346  &  45.8845101  & 20100803  & 1280.6 $\pm$ 4.9  & -  & 500.7 $\pm$ 2.5  & 409.8 $\pm$ 3.0 \\
\hline
\end{tabular}
\end{center}
$^*${Observations marked with * are VLA observations done at K (22\,GHz) instead of Ka (33\,GHz).}
\end{table*}

\begin{table*}[t]
\caption{Matches of our VLA/JVLA sources with sources in the ERCSC catalog} \label{table_matches} 
\begin{center}
\begin{tabular}{ccll}
\hline\hline
J2000 Name & \Planck Name & (J)VLA Dateobs & Planck Dateobs \\
\hline
J0006-0623  &  PLCKERC030 G093.49-66.62  &  20100702  &  20091221 \\
J0108+0135  &  PLCKERC030 G131.82-60.99  &  20100723  &  20100112 \\
J0125-0005  &  PLCKERC030 G141.26-61.77  &  20100803  &  20100115,20100116 \\
J0137+3309  &  PLCKERC030 G133.94-28.63  &  20100702,20100723,20100803, &  20100130,20100131 \\
 & & 20100907,20100920,20100924, &  \\
 & & 20101020,20101109,20101127, &  \\
 & & 20101120,20091115,20090827  &  \\
J0217+7349  &  PLCKERC030 G128.95+11.97  &  20100920,20100924,20090827  &  20090911,20090912,20100219, \\
 & &  & 20100220 \\
J0228+6721  &  PLCKERC030 G132.15+06.22  &  20100920  &  20090906,20100217,20100218 \\
J0319+4130  &  PLCKERC030 G150.59-13.25  &  20100907  &  20090829,20100218,20100219 \\
J0336+3218  &  PLCKERC030 G159.01-18.79  &  20100907  &  20090828,20090829,20100220 \\
J0359+5057  &  PLCKERC030 G150.36-01.65  &  20100920  &  20090907,20090908,20100225 \\
J0418+3801  &  PLCKERC030 G161.66-08.78  &  20100920  &  20090906,20090907,20100227, \\
 & &  & 20100228 \\
J0423-0120  &  PLCKERC030 G195.26-33.13  &  20100907,20091103,20091103  &  20090828,20090829,20100226, \\
 & &  & 20100227 \\
J0433+0521  &  PLCKERC030 G190.40-27.42  &  20100907  &  20090901,20100301 \\
J0449+1121  &  PLCKERC030 G187.35-20.74  &  20100907,20100920  &  20090906,20100303 \\
J0501-0159  &  PLCKERC030 G201.45-25.29  &  20100907  &  20090905,20090906,20100305 \\
J0530+1331  &  PLCKERC030 G191.42-11.03  &  20100920,20100924  &  20090914,20090915,20100310, \\
 & &  & 20100311 \\
J0555+3948  &  PLCKERC030 G171.66+07.28  &  20100301,20100924,20101018, &  20090922,20100313,20100314 \\
 & & 20101130  &  \\
J0607-0834  &  PLCKERC030 G215.75-13.51  &  20100924  &  20090919,20090920,20100322, \\
 & &  & 20100323 \\
J0646+4451  &  PLCKERC030 G171.09+17.92  &  20100924  &  20090930,20091001,20100322 \\
J0721+7120  &  PLCKERC030 G143.98+28.03  &  20101018  &  20091005,20100317,20100318 \\
J0725-0054  &  PLCKERC030 G217.69+07.21  &  20101018,20091103  &  20091009,20091010,20100407 \\
J0738+1742  &  PLCKERC030 G201.84+18.14  &  20101018  &  20091010,20091011,20100406 \\
J0739+0137  &  PLCKERC030 G216.97+11.36  &  20101018,20091022  &  20091012,20091013,20100409, \\
 & &  & 20100410 \\
J0745-0044  &  PLCKERC030 G219.91+11.78  &  20091103  &  20091014,20091015,20100411, \\
 & &  & 20100412 \\
J0750+1231  &  PLCKERC030 G208.18+18.75  &  20091103,20101018  &  20091014,20100410 \\
J0757+0956  &  PLCKERC030 G211.33+19.05  &  20101018,20091022  &  20091016,20100411,20100412 \\
J0825+0309  &  PLCKERC030 G221.26+22.36  &  20101108,20091022  &  20091024,20091025,20100419, \\
 & &  & 20100420 \\
J0830+2410  &  PLCKERC030 G200.06+31.89  &  20091022  &  20091021,20100415,20100416 \\
J0841+7053  &  PLCKERC030 G143.55+34.41  &  20101018,20091022  &  20091012,20100326,20100327 \\
J0854+2006  &  PLCKERC030 G206.78+35.81  &  20101108  &  20091027,20100421,20100422 \\
J0909+4253  &  PLCKERC030 G178.32+42.86  &  20091118  &  20091024,20100417 \\
J0920+4441  &  PLCKERC030 G175.71+44.81  &  20091118,20091118,20101108  &  20091025,20100418,20100419 \\
J0927+3902  &  PLCKERC030 G183.72+46.16  &  20091118,20101108  &  20091028,20100422,20100423 \\
J0948+4039  &  PLCKERC030 G181.02+50.31  &  20091118,20101108  &  20091031,20100426,20100427 \\
J0956+2515  &  PLCKERC030 G205.49+50.96  &  20101108,20101130  &  20091108,20100505 \\
J1043+2408  &  PLCKERC030 G211.59+61.00  &  20101130  &  20091119,20091120,20100519 \\
J1058+0133  &  PLCKERC030 G251.59+52.70  &  20100103  &  20091207,20091208,20100601, \\
 & &  & 20100602 \\
J1130+3815  &  PLCKERC030 G174.47+69.79  &  20101130,20100103  &  20091121,20100527 \\
J1153+4931  &  PLCKERC030 G145.58+64.96  &  20100103  &  20091116,20091117,20100525, \\
 & &  & 20100526 \\
J1153+8043  &  PLCKERC030 G125.75+35.85  &  20091103  &  20091016,20091017,20100321, \\
 & &  & 20100322 \\
J1159+2914  &  PLCKERC030 G199.42+78.39  &  20100103,20091103  &  20091204,20091205 \\
J1222+0413  &  PLCKERC030 G284.63+66.05  &  20100704,20100709  &  20091229,20091230 \\
J1224+2122  &  PLCKERC030 G255.00+81.65  &  20100103  &  20091218,20091219 \\
J1229+0203  &  PLCKERC030 G290.02+64.36  &  20100704,20100709  &  20100102 \\
J1230+1223  &  PLCKERC030 G283.75+74.54  &  20100704,20100709  &  20091226,20091227 \\
J1310+3220  &  PLCKERC030 G085.86+83.31  &  20100709,20100103  &  20091221,20091222 \\
J1327+2210  &  PLCKERC030 G003.62+80.50  &  20100709  &  20100105,20100106 \\
J1331+3030  &  PLCKERC030 G056.70+80.65  &  20100704,20100709,20100915, &  20091230,20091231 \\
 & & 20101108,20091118,20100103  &  \\
J1419+5423  &  PLCKERC030 G098.28+58.29  &  20100704,20100709,20100103  &  20091206,20091207 \\
\hline
\end{tabular}
\end{center}
\end{table*}

\setcounter{table}{3}

\begin{table*}[t]
\caption{ \it{Continued}} 
\begin{center}
\begin{tabular}{ccll}
\hline\hline
J2000 Name & \Planck Name & EVLA Dateobs & Planck Dateobs \\
\hline
J1642+6856  &  PLCKERC030 G100.69+36.62  &  20100704,20100709  &  20091106,20091107,20091108, \\
 & &  & 20100222,20100223,20100224, \\
 & &  & 20100225,20100226,20100409, \\
 & &  & 20100410,20100411,20100412, \\
 & &  & 20100413,20100414 \\
J1743-0350  &  PLCKERC030 G021.60+13.15  &  20100915  &  20090914,20100317 \\
J1751+0939  &  PLCKERC030 G034.92+17.63  &  20100915  &  20090913,20090914,20100321 \\
J1800+3848  &  PLCKERC030 G065.15+26.02  &  20100915  &  20090907,20100329,20100330 \\
J1800+7828  &  PLCKERC030 G110.05+29.07  &  20101020  &  20091011,20091012,20100218, \\
 & &  & 20100219 \\
J1806+6949  &  PLCKERC030 G100.12+29.18  &  20100704,20100709,20101109, &  20091022,20091023,20100129, \\
 & & 20101127  & 20100130,20100131,20100511, \\
 & &  & 20100512,20100513,20100514 \\
J1824+5651  &  PLCKERC030 G085.72+26.08  &  20100915  &  20090827,20090828,20090829, \\
 & &  & 20100417,20100418 \\
J1829+4844  &  PLCKERC030 G077.20+23.49  &  20100915  &  20090910,20090911,20090912, \\
 & &  & 20100411,20100412 \\
J1849+6705  &  PLCKERC030 G097.46+25.04  &  20100723,20100803,20100915, &  20091012,20091013,20091014, \\
 & & 20101020,20101109  & 20100117,20100118,20100119, \\
 & &  & 20100513,20100514 \\
J1927+7358  &  PLCKERC030 G105.63+23.54  &  20101020  &  20090928,20090929,20090930, \\
 & &  & 20100202,20100203 \\
J1955+5131  &  PLCKERC030 G085.28+11.76  &  20101020,20101127,20101120  &  20091203,20091204,20091205, \\
 & &  & 20100506,20100507 \\
J2005+7752  &  PLCKERC030 G110.45+22.73  &  20101020  &  20090925,20090926,20100209, \\
 & &  & 20100210 \\
J2022+6136  &  PLCKERC030 G096.08+13.77  &  20100702,20100723,20100803  &  20100106,20100107,20100526, \\
 & &  & 20100527,20100528 \\
J2123+0535  &  PLCKERC030 G058.03-30.09  &  20101109,20091115  &  20091110,20091111,20100509 \\
J2134-0153  &  PLCKERC030 G052.38-36.49  &  20091115,20091103  &  20091110,20100509,20100510 \\
J2136+0041  &  PLCKERC030 G055.47-35.57  &  20101109,20091115  &  20091112,20100511 \\
J2139+1423  &  PLCKERC030 G068.51-27.50  &  20101127,20101120,20091115  &  20091120,20091121,20100516 \\
J2148+0657  &  PLCKERC030 G063.61-34.11  &  20091115  &  20091118,20091119,20100516 \\
J2202+4216  &  PLCKERC030 G092.62-10.44  &  20091103  &  20091225,20091226,20100606, \\
 & &  & 20100607 \\
J2203+1725  &  PLCKERC030 G075.68-29.62  &  20101127,20101120,20091103  &  20091201,20091202,20100524, \\
 & &  & 20100525 \\
J2203+3145  &  PLCKERC030 G085.95-18.77  &  20100702,20091103  &  20091214,20091215,20100531 \\
J2218-0335  &  PLCKERC030 G059.05-46.63  &  20091115  &  20091121,20091122,20100522 \\
J2225-0457  &  PLCKERC030 G058.96-48.81  &  20091115  &  20091123,20100523,20100524 \\
J2229-0832  &  PLCKERC030 G055.15-51.70  &  20091115  &  20091122,20100523,20100524 \\
J2232+1143  &  PLCKERC030 G077.45-38.54  &  20091103  &  20091206,20100601 \\
J2236+2828  &  PLCKERC030 G090.11-25.64  &  20100702,20091103  &  20091220,20091221 \\
J2253+1608  &  PLCKERC030 G086.12-38.18  &  20100702,20091103  &  20091215,20091216 \\
J2327+0940  &  PLCKERC030 G091.17-47.99  &  20100702,20100723,20100803  &  20091220,20091221 \\
\hline
\end{tabular}
\end{center}
\end{table*}


\begin{table*}[ht]
\begin{center}
\caption{The relative frequency of different SED types, as defined in Figure\,\ref{color_color}. Due to incomplete spectral coverage, this includes only 60 of our sources. The distribution of SED types for the PACO Bright and Faint samples are given for comparison \citep{PACO-proj, PACO-proj-faint}.} \label{table_classes}
\begin{tabular}{cccc}
\hline\hline
SED Type & JVLA Sample & PACO Bright Sample & PACO Faint Sample\\
\hline
Flat & 42\% & 10\% & 5\% \\
Steep & 10\% & 4\% & 13\% \\
Down Turning & 23\% & 66\% & 65\% \\
Up Turning & 7\% & 0\% & 0\% \\
Inverted & 3\% & 1\% & 0\% \\
Self Absorbed & 8\% & 5\% & 5\% \\
Peaked & 7\% & 15\% & 11\% \\
\hline
\end{tabular}
\end{center}
\end{table*}

\begin{table*}[ht!]
\begin{center}
\caption{Median variability indices\label{table_var}. The values in parenthesis exclude the two calibrators 3C48 and 3C286.}
\begin{tabular}{ccccc}
\hline\hline
Band & 0-60days & 60-260days & 290-430days & 1-7years \\
\hline
5\,GHz &  1.0(1.1) & 4.0(4.3)& 3.0(3.3) & -- \\
33\,GHz &  2.2(2.9) & 4.0(5.3) &  10.8(14.4) & 15.6()  \\
\hline
\end{tabular}
\end{center}
\end{table*}

\end{document}